\documentclass{article}
\usepackage{latexsym}
\usepackage{amsmath}
\usepackage{amsthm}
\usepackage{amsfonts}
\usepackage{amsxtra}
\usepackage{hyperref}
\allowdisplaybreaks
\title{Rotating frames and gauge invariance in three-dimensional
  many-body quantum systems}
\author{Antonio O.\ Bouzas\thanks{abouzas@mda.cinvestav.mx}\
  \mbox{ }and
  Jos\'e M\'endez Gamboa\thanks{jmendez@mda.cinvestav.mx}
  \\\small
  Departamento de F\'{\i}sica Aplicada, CINVESTAV-IPN
  \\[-0.5ex]\small Carretera Antigua a Progreso Km.\ 6,
  Apdo.\ Postal 73 ``Cordemex''\\[-0.5ex]\small M\'erida 97310,
  Yucat\'an, M\'exico}
\date{February 27, 2004 (Revised May 12, 2004)}
\newcommand{\Ll}{\ensuremath{\mathcal{L}}}
\newcommand{\Llrt}{\ensuremath{\Ll_\mathrm{rt}}}
\newcommand{\Lln}{\ensuremath{\Ll_N}}
\newcommand{\Llcm}{\ensuremath{\Ll_\mathrm{cm}}}
\newcommand{\Lrt}{\ensuremath{\boldsymbol{s}}}
\newcommand{\V}{\ensuremath{\mathcal{V}}}
\newcommand{\I}{\ensuremath{\mathcal{I}}}
\renewcommand{\H}{\ensuremath{\mathcal{H}}}
\newcommand{\Hrt}{\ensuremath{\H_\mathrm{rt}}}
\newcommand{\Hn}{\ensuremath{\H_N}}
\newcommand{\Hnr}{\ensuremath{\wh{\H}_N}}
\newcommand{\Hnt}{\ensuremath{\wt{\H}_N}}
\newcommand{\Hntr}{\ensuremath{\wth{\H}_N}}

\newcommand{\U}{\ensuremath{\mathcal{U}}}
\newcommand{\mat}[1]{\ensuremath{\boldsymbol{#1}}}
\newcommand{\matd}[1]{\ensuremath{\boldsymbol{\dot{#1}}}}
\newcommand{\matdd}[1]{\ensuremath{\boldsymbol{\ddot{#1}}}}
\newcommand{\ver}[1]{\ensuremath{\boldsymbol{\widehat{#1}}}}
\newcommand{\verd}[1]{\boldsymbol{\dot{\widehat{#1}}}}

\newcommand{\wt}[1]{\widetilde{#1}}
\newcommand{\wh}[1]{\widehat{#1}}
\newcommand{\wth}[1]{\widehat{\widetilde{#1\rule{1pt}{0pt}}}}
\renewcommand{\S}{\ensuremath{\mathfrak{S}}}

\newcommand{\Q}{\ensuremath{\mathfrak{Q}}}

\newcommand{\R}{\ensuremath{\mathfrak{R}}}

\newcommand{\C}{\ensuremath{\mathfrak{C}}}
\newcommand{\G}{\ensuremath{\mathfrak{G}}}
\newcommand{\J}{\ensuremath{\mathcal{J}}}
\newcommand{\N}{\ensuremath{\mathcal{N}}}
\newcommand{\F}{\ensuremath{\mathcal{F}}}
\newcommand{\D}{\ensuremath{\mathcal{D}}}
\newcommand{\vrcm}{\ensuremath{\mat{r}_\mathrm{cm}}}
\newcommand{\vdrcm}{\ensuremath{\matd{r}_\mathrm{cm}}}
\newcommand{\rcm}{\ensuremath{r_\mathrm{cm}}}
\newcommand{\vpcm}{\ensuremath{\mat{p}_\mathrm{cm}}}
\newcommand{\pcm}{\ensuremath{p_\mathrm{cm}}}
\newcommand{\vlcm}{\ensuremath{\mat{l}_\mathrm{cm}}}
\newcommand{\lcm}{\ensuremath{l_\mathrm{cm}}}
\newcommand{\syl}{\scriptstyle}
\newcommand{\lgt}{\begin{smallmatrix} >\\<\end{smallmatrix}}
\renewcommand{\O}{\mathcal{O}}
\begin{document}
\maketitle
\begin{abstract}
  We study the quantization of many-body systems in three dimensions
  in rotating coordinate frames using a gauge invariant formulation of
  the dynamics.  We consider reference frames defined by linear gauge
  conditions, and discuss their Gribov ambiguities and commutator
  algebra.  We construct the momentum operators, inner-product and
  Hamiltonian in those gauges, for systems with and without
  translation invariance.  The analogy with the quantization of
  non-Abelian Yang-Mills theories in non-covariant gauges is
  emphasized.  Our results are applied to quasi-rigid systems in the
  Eckart frame.
\end{abstract}
\section{Introduction}
\label{sec:intro}

The problem of quantizing a many-body mechanical system in a rotating
reference frame is of interest both by itself and for its possible
applications to specific problems in, e.g., molecular and nuclear
physics.  In this paper we study the quantization of many-body sytems
in three dimensions in rotating coordinate frames, using a
gauge-invariant formulation.  The two-dimensional case was considered
in a previous paper \cite{bou}, in which the method was developed in
detail and a close parallel with the quantization of electrodynamics
in non-covariant gauges established.  The main lines of the method are
the same in both cases.  Due to the non-Abelianity of the rotation
group in three dimensions, however, the technical treatment of the
systems considered here is considerably different from the planar
case, the differences being already apparent at the Lagrangian level
as discussed in the following section.

We consider systems of $N$ spinless particles interacting through
central potentials.  Since the underlying dynamics are rotationally
symmetric, the coordinate transformation from a space-fixed reference
frame to a rotating one with the same origin is a time-dependent
symmetry transformation, or gauge transformation.  If the dynamics are
described in terms of a gauge-invariant action, since we know how to
quantize a mechanical system in a space-fixed coordinate frame, we can
perform a gauge transformation in order to obtain the quantum theory
in a rotating frame.  Gauge invariance implies that both theories are
physically equivalent.

Rotating frames are often defined implicitly, by restrictions on the
trajectories of the system in that frame.  In the gauge-invariant
approach to the quantization in rotating frames, such restrictions are
incorporated into the theory as gauge conditions.  The action is then
given in terms of degrees of freedom that are not independent, but
must satisfy certain functional relations.  This situation is familiar
from the theory of gauge fields \cite{lee,lnz}, where the vector
potential $\mat{A}(t,\mat{x})$ may be required to satisfy such
relations as $\mat{\nabla}\cdot\mat{A}=0$ (Coulomb gauge), or
$\mat{n}\cdot\mat{A}=0$ (axial gauge), at all times $t$.  In this
paper we consider only gauge conditions depending linearly on the
particles coordinates, which are most useful in practical applications
involving perturbative expansions.  We do not consider quadratic gauge
conditions, in particular, because we expect the formalism in those
gauges to be considerably more complicated, in view of the results of
\cite{bou} in the simpler two-dimensional, Abelian case.  Furthermore,
the quadratic gauge conditions most common in the literature
\cite{bs2,vil} are those defining the instantaneous principal axes
frame, in which the total angular momentum of the system is strongly
coupled to the other degrees of freedom through the inertia tensor and
the Coriolis terms.  Without a strong physical motivation for
quadratic gauges, we have no reason to pursue that technically more
involved approach here.  As discussed in \cite{bou}, however, there is
no problem of principle to deal with those and other kinds of gauge
conditions within the formalism espoused in this paper.

We closely follow the approach to non-Abelian Yang-Mills theories in
non-covariant gauges of \cite{lee,chr}, stressing throughout the paper
the strict formal similarity between our results for many-body systems
and the corresponding ones in \cite{chr} for Yang-Mills theories.  Our
goals are both to illustrate the formalism of gauge theories in the
more familiar context of mechanical systems, and to apply the
gauge-theoretical techniques to the quantization of three-dimensional
$N$-body sytems in rotating frames.  Previous treatments of the latter
problem within a gauge-invariant approach have been given in
\cite{bs2} and references therein. A gauge theory of rotations and
internal motions of deformable bodies, including classical and quantum
$N$-body systems, is developed in \cite{ljn} (see also \cite{mrm})
from a point of view different from ours.  Non-gauge-invariant
treatments can be found, e.g., in \cite{vil} in the context of nuclear
physics, and in \cite{gai,bie} in molecular physics.

The outline of the paper is as follows.  In section \ref{sec:model} we
describe the class of systems to be considered throughout the paper,
and their formulation in terms of a Lagrangian invariant under
time-dependent rotations.  Their quantization in a space-fixed frame
is given, and shown to be equivalent to the non-gauge-invariant
formulation.  The central results of the paper are given in section
\ref{sec:linear}, where we consider the quantization in rotating
frames defined by linear gauge conditions.  We discuss in detail the
commutator algebra for both linear and angular momentum operators, and
give explicit realizations of that algebra in terms of differential
operators.  Those operators are used to construct the Hamiltonian in
terms of position vectors referred to the rotating frame, and their
conjugate momenta.  The elimination of orientational degrees of
freedom from the formalism is subsequently carried out, and the
resulting Hamiltonian and its Weyl-ordered form and related quantum
potential are obtained.  As emphasized throughout, by describing a
many-body system from a rotating frame defined by gauge conditions, we
are introducing orthogonal curvilinear coordinates in configuration
space.  The singularities of those coordinates occur at the Gribov
horizons where the gauge conditions become ambiguous.  Gribov
ambiguities \cite{grb,chr,lnz,bou} are discussed in detail in relation
to the construction of the inner product in the reduced state-space of
the system.

In section \ref{sec:cmm} we extend the results of section
\ref{sec:linear} to translation-invariant systems.  We show how the
gauge-invariant approach can be used to describe a mechanical system
in a reference frame in an arbitrary state of rotation and
translation.  In particular, we obtain explicit results for the
quantization of $N$-body systems in rotating frames with origin at the
center of mass.  Those results are then applied to quasi-rigid systems
in the Eckart frame \cite{ek2,gai,bou} in section \ref{sec:quasi},
where two simple three- and four-body examples are briefly discussed.
In section \ref{sec:finrem} we give our final remarks.  Some
complementary material is gathered in the appendices.  

\section{\mat{N}-particle system}
\label{sec:model}

We consider a system of $N$ spinless particles with central
interactions in three dimensions, described by the Lagrangian
\begin{equation}\label{eq:rlag}
\begin{gathered}
  \Ll = \Lln + \Llrt~,
  \qquad
  \Lln = \frac{1}{2} \sum_{\alpha=1}^N m_\alpha \matd{r}^2_\alpha
  - \V \\
  \V = \sum_{\alpha < \beta =1}^N V_{\alpha\beta}(|\mat{r}_\alpha -
  \mat{r}_\beta |) + 
  \sum_{\alpha=1}^N U(r_\alpha)~,
  \qquad 
  \Llrt = \frac{\I}{2}
  \left(\ver{e}\wedge\verd{e}\right)^2 =
  \frac{\I}{2} \verd{e}^2~,
  \quad
  \ver{e}\cdot\ver{e} = 1~.
\end{gathered}
\end{equation}
The potential energy \V\ is chosen for concreteness to include only
one- and two-body interactions.  Our results do not depend on that
fact and apply equally well to more general rotationally invariant
potentials.  If the one-body potential $U = 0$, \Ll\ is invariant
under the group of Euclidean motions of three-dimensional space. In
this and the following sections we consider $U \neq 0$ and focus on
the non-Abelian group of three-dimensional rotations, deferring the
discussion of translation invariance until section \ref{sec:cmm}.
Besides the kinetic and potential energy for the $N$-particle system
\Ll\ also contains the Lagrangian \Llrt\ for a free rigid rotator,
described by a unit vector $\ver{e}$.  This rotator is not coupled to
the particle system, so it does not affect its dynamical evolution.
\Llrt\ can be made to vanish by letting the rotator's inertia moment
$\I\rightarrow\infty$ while keeping constant the magnitude of its
angular momentum $\Lrt = \I \ver{e} \wedge \verd{e}$.  Whereas \Llrt\ 
does not play any r\^ole in \Ll\ as given in (\ref{eq:rlag}), it will
serve as a source of angular momentum for the particle system in the
gauge-invariant formulation to which we now turn.

\Ll\ is invariant under time-independent rotations of the coordinate
frame.  In order to make \Ll\ invariant under changes of arbitrarily
rotating coordinate frames we apply the Yang-Mills construction
\cite{yml} to (\ref{eq:rlag}).  We introduce a $3\times3$ real
antisymmetric matrix \mat{\xi}, thus adding three new degrees of
freedom to the system, and postulate the following transformation laws
under rotations of the coordinate frame,
\begin{equation}\label{eq:trans}
  \mat{r}_\alpha^\prime = \mat{U}\mat{r}_\alpha~,
  \qquad
  \ver{e}\,^\prime = \mat{U}\ver{e}~,
  \qquad
  \mat{\xi}^\prime = \mat{U} \mat{\xi} \mat{U}^\dagger + \matd{U}
  \mat{U}^\dagger~, 
\end{equation}
with \mat{U}\ a time-dependent real orthogonal matrix and
$\mat{U}^\dagger$ its transpose.  These are the gauge transformations
of the system.  We define the covariant time-derivative $D_t
\mat{r}_\alpha \equiv \matd{r}_\alpha - \mat{\xi} \mat{r}_\alpha$,
which transforms like a vector under gauge transformations, $(D_t
\mat{r}_\alpha)^\prime = \mat{U}(D_t \mat{r}_\alpha)$.  Analogously,
$D_t \ver{e} = \verd{e} - \mat{\xi} \ver{e}$.  We can, equivalently,
use instead of the matrix \mat{\xi}\ the axial vector $\wt{\mat{\xi}}$
dual to \mat{\xi}, $\xi_{ik} = \varepsilon_{ijk}\wt\xi_{j}$, whose
gauge transformations can be derived from (\ref{eq:trans}).  In terms
of $\wt{\mat{\xi}}$ covariant derivatives take the form $D_t
\mat{r}_\alpha = \matd{r}_\alpha - \wt{\mat{\xi}}
\wedge\mat{r}_\alpha$.  Substituting time derivatives in
(\ref{eq:rlag}) by covariant derivatives we obtain a Lagrangian
invariant under time-dependent rotations of the coordinate frame.
Explicitly, we write,
\begin{equation}\label{eq:glag}
\begin{split}
  \Ll & = \Lln + \Llrt~,\\
  \Lln & = \frac{1}{2} \sum_{\alpha=1}^N m_\alpha (D_t\mat{r}_\alpha)^2
  - \V 
    = \frac{1}{2} \sum_{\alpha=1}^N m_\alpha
  \matd{r}_\alpha^2 + \frac{1}{2} \sum_{\alpha=1}^N m_\alpha
  \left(\wt{\mat{\xi}} \wedge \mat{r}_\alpha \right)^2 -
  \wt{\mat{\xi}} \cdot \sum_{\alpha=1}^N m_\alpha \left(
    \mat{r}_\alpha
    \wedge \matd{r}_\alpha \right) - \V~, \\
  \Llrt & = \frac{\I}{2} \left(\ver{e}\wedge (D_t\ver{e})\right)^2 =
  \frac{\I}{2} \left(D_t\ver{e}\right)^2 = \frac{\I}{2} \verd{e}^2 +
  \frac{\I}{2} \left(\wt{\mat{\xi}} \wedge \ver{e}\right)^2 -
  \I\wt{\mat{\xi}} \cdot \left(\ver{e} \wedge
    \verd{e}\right)~,
\end{split}
\end{equation}
where the potential energy \V\ is defined in (\ref{eq:rlag}).  \Ll\ is
exactly invariant under the gauge transformations (\ref{eq:trans})
and, in fact, \Lln\ and \Llrt\ are separately invariant under
(\ref{eq:trans}).\footnote{Notice that, unlike the two-dimensional
  (Abelian) case \cite{bou}, we cannot add external-source terms
  to \Ll\ without breaking gauge invariance, so we have to incorporate
  the source into the theory as a dynamical degree of freedom.  That
  is the motivation for including \Llrt\ in \Ll.  We stress
  here that \Llrt\ is gauge invariant and therefore it is not a 
  gauge fixing term.}

\Lln\ in (\ref{eq:glag}) has the form of a Lagrangian for an
$N$-particle system described from a coordinate frame rotating with
angular velocity $-\wt{\mat{\xi}}$ with respect to the laboratory
frame \cite{lan}.  Notice, however, that $\wt{\mat{\xi}}$ is a
dynamical variable describing the coupling of the particles and the
rotator to the inertial forces.  The equations of motion for
$\mat{r}_\alpha$, \ver{e}, and $\wt{\mat{\xi}}$ derived from \Ll\ are
\begin{subequations}\label{eq:eqm}
  \begin{gather}
    m_\alpha D_t D_t \mat{r}_\alpha + \mat{\nabla}_\alpha\V = m_\alpha
    \matdd{r}_\alpha - 2m_\alpha \wt{\mat{\xi}} \wedge \matd{r}_\alpha
    - m_\alpha \matd{\wt{\xi}} \wedge \mat{r}_\alpha - m_\alpha
    \wt{\mat{\xi}}\wedge \left(\mat{r}_\alpha \wedge
      \wt{\mat{\xi}}\right) + \mat{\nabla}_\alpha\V =
    0~, \label{eq:eqma}\\
    D_t D_t \ver{e} + \left(D_t \ver{e}\right)^2 \ver{e} =0 \quad
    \text{with} \quad \ver{e} \cdot \ver{e} =1~,
    \label{eq:eqmb}\\
    -\frac{\partial \Ll}{\partial \mat{\wt{\xi}}} = 
    \sum_{\alpha=1}^N m_\alpha \mat{r}_\alpha \wedge \left( D_t
    \mat{r}_\alpha \right) + \I \ver{e} \wedge \left(D_t\ver{e}\right)
    = 0~.  \label{eq:eqmc}
  \end{gather}
\end{subequations}
In the equation of motion (\ref{eq:eqma}) for $\mat{r}_\alpha$ the
terms due to the Coriolis, azimuthal and centrifugal forces \cite{mek}
are apparent.  As a consequence of the rotational invariance of \Ll\ 
the total angular momentum of the system is conserved,
$d\mat{j}/dt=0$ with,
\begin{equation}
  \label{eq:angmom}
  \mat{j} = \mat{l} + \Lrt~, \qquad
  \mat{l} = \sum_{\alpha =1}^N m_\alpha \mat{r}_\alpha \wedge
  (D_t\mat{r}_\alpha)~, \qquad
  \Lrt = \I \ver{e} \wedge (D_t\ver{e})~,
\end{equation}
Clearly, the vector \mat{j}\ can be time-independent in every rotating
reference frame only if it vanishes.  This is expressed by
(\ref{eq:eqmc}), which can be rewritten as $\mat{j} = 0$.  Since in
general \Ll\ is not invariant under separate rotations of
$\{\mat{r}_\alpha\}$ and \ver{e}, \mat{l}\ and \Lrt\ are not
separately conserved.  Rather, from (\ref{eq:eqma}) and
(\ref{eq:eqmb}) they are seen to be covariantly conserved,
\begin{equation}
  \label{eq:covcon}
  D_t\mat{l} = 0~, \qquad D_t\Lrt = 0~,
\end{equation}
with $D_t\mat{l} = \matd{l} - \mat{\wt{\xi}} \wedge \mat{l}$.  {F}rom
(\ref{eq:covcon}), the magnitudes of \mat{l}\ and \Lrt\ are conserved
and frame-independent, but their directions in space are
time-dependent.  Only in the lab frame (in which $\mat{\xi} = 0$ and
$D_t = d/dt$, as discussed below) are \mat{l}\ and \Lrt\ conserved.

Since the system is gauge invariant we can fix the gauge by imposing a
set of conditions of the form\footnote{The letters $a,b,c,d$ are used
  to index quantities which are not necessarily tensor components,
  such as $\G_a$.  Summation over those indices and their ranges of
  variation are always explicitly indicated.  We only use the
  summation convention for tensor indices, which are denoted by latin
  letters $i,j,k,l,\ldots$ and always run from 1 to 3.}
$\G_a(\{\mat{r}_\alpha\},\mat{\xi},\ver{e}) = 0$, $a=1,2,3$, which is
equivalent to selecting a rotating frame in which the trajectory of
the system $(\{\mat{r}_\alpha(t)\},\mat{\xi}(t),\ver{e}(t))$ in
configuration space is constrained by the relations
$\G_a(\{\mat{r}_\alpha(t)\},\mat{\xi}(t),\ver{e}(t)) = 0$.  The
functions\footnote{In general, $\G_a$ are functionals of the
  trajectory $(\{\mat{r}_\alpha(t)\},\mat{\xi}(t),\ver{e}(t))$.}
$\G_a$, $a=1,2,3$, can be chosen arbitrarily, as long as any
trajectory
$(\{\mat{r}_\alpha^\prime\},\mat{\xi}^\prime,\ver{e}\,^\prime)$ can be
transformed into a new one $(\{\mat{r}_\alpha\},\mat{\xi},\ver{e})$
satisfying $\G_a=0$.  The new trajectory must be unique, in the sense
that no other trajectory obtained from
$(\{\mat{r}_\alpha^\prime\},\mat{\xi}^\prime,\ver{e}\,^\prime)$ by a
gauge transformation satisfies the gauge conditions. Otherwise, the
gauge is said to be ambiguous \cite{grb}.  Supplementary conditions
must then be imposed to fix the ambiguity.

\subsection{The laboratory frame}
\label{sec:labo}

Given any trajectory of the system
$(\{\mat{r}_\alpha(t)\},\mat{\xi}(t),\ver{e}(t))$ by means of a gauge
transformation we can obtain a physically equivalent trajectory with
$\mat{\xi}^\prime = \mat{U} \mat{\xi} \mat{U}^\dagger + \matd{U}
\mat{U}^\dagger = 0$.  Indeed, given the antisymmetric matrix-valued
function of time $\mat{\xi}(t)$, there is always an orthogonal matrix
$\mat{U}(t)$ satisfying $\mat{U}^\dagger \matd{U} = -\mat{\xi}$.  The
condition $\mat{\xi} = 0$ is then admissible as a choice of gauge for
the system, which corresponds to selecting a non-rotating coordinate
frame referred to as the ``laboratory frame.''  

We denote dynamical quantities in the laboratory frame by lower-case
symbols, except for the Lagrangian and Hamiltonian.  In this gauge the
Lagrangian (\ref{eq:glag}) reduces to (\ref{eq:rlag}).  \Ll\ is
invariant under separate rotations of $\{\mat{r}_\alpha\}$ and
$\ver{e}$, leading to the separate conservation of the angular momenta
$\mat{l} = \sum_{\alpha=1}^N m_\alpha \mat{r}_\alpha \wedge
\matd{r}_\alpha$ of the system of particles and $\mat{s} = \I \ver{e}
\wedge \verd{e}$ of the rigid rotator.  The equation of motion
(\ref{eq:eqmc}) for \mat{\xi}, which cannot be obtained from
(\ref{eq:rlag}), must be imposed on the system as a constraint
\cite{lee}, $\mat{j} \equiv \mat{l} + \mat{s} = 0$.  In the
Hamiltonian formulation in this gauge, this is a primary first-class
constraint \cite{dir}, not leading to further secondary ones.

The quantization in the gauge $\mat{\xi}=0$ is canonical.  In units
such that $\hbar=1$ we have,
\begin{equation}
  \label{eq:xi0}
  \begin{gathered}
    \H = \Hn + \Hrt, \qquad
    \Hn = \sum_{\alpha=1}^N \frac{1}{2 m_\alpha} \mat{p}_\alpha^2 +
    \V~, \qquad
    \Hrt = \frac{1}{2\I} \mat{s}^2~,\\
    \left[ r_{\alpha i}, p_{\beta j} \right]  = i \delta_{\alpha\beta}
    \delta_{ij}~,      \qquad
    \mat{p}_\alpha = \frac{1}{i} \mat{\nabla}_\alpha, \qquad
    \left[ s_{i}, s_{j} \right]  = i \varepsilon_{ijk} s_k~, \\
    \langle\phi | \psi\rangle = \int \prod_{\beta=1}^N d^3r_\beta
    d^2\ver{e}\, \phi^*\left(\{\mat{r}_\alpha\},\ver{e}\right)
    \psi\left(\{\mat{r}_\alpha\},\ver{e}\right)~,
  \end{gathered}
\end{equation}
with the commutators among $\mat{r}_\alpha$ and $\mat{p}_\alpha$ not
shown in (\ref{eq:xi0}) all vanishing.  The first-class constraint is
imposed on the state space \cite{dir}, $\mat{j} |\psi\rangle = 0$.
Since both \mat{l}\ and \mat{s}\ are constants of motion, this
constraint is clearly consistent with the dynamics.  We see that the
quantized theory in the $\mat{\xi}=0$ gauge is completely analogous to
Yang-Mills theories in the temporal gauge \cite{chr,grv,tdr}.  The
constraint fixing the value of \mat{j}, in particular, is the
equivalent of the non-Abelian Gauss law.  In the constraint equation
\mat{s}\ plays the same r\^ole as the fermion color current in Gauss
law.

If in (\ref{eq:xi0}) we let $\I \rightarrow \infty$ with $\mat{s}^2$
fixed, $\Hrt\rightarrow 0$ and the rigid rotator drops from the
Hamiltonian, entering the dynamics only through the constant value of
$\mat{s}$ in the constraint.  Thus, (\ref{eq:xi0}) describes in that
limit an $N$-body system with interaction potential \V\ in the sector
of fixed angular momentum $\mat{l} = -\mat{s}$ (i.e., the null
eigenspace of $(\mat{l}+\mat{s})^2$).  Due to gauge invariance, the
same must be true in any other gauge.

\section{Linear gauge conditions}
\label{sec:linear}

In order to fix a reference frame we need to impose three gauge
conditions.  The simplest gauge conditions involving the coordinates
of the particles depend linearly on $\{\mat{r}_\alpha\}$, and do not
involve \mat{\xi}\ or \ver{e}.  As discussed below, linear gauges are
relevant in the context of perturbative or semiclassical expansions.
The general form of the linear gauge conditions is,
\begin{equation}
  \label{eq:linear}
  \S_a(\{\mat{r}_\alpha\}) \equiv \sum_{\alpha=1}^N m_\alpha
  \Gamma_{a\alpha j} r_{\alpha j} = 0~,
  \qquad
  a=1,2,3,
\end{equation}
with $\Gamma_{a\alpha j}$ a set of $9N$ constants defining the
functions $\S_a$.  We denote dynamical quantities in this gauge by
capital letters, in particular the position vectors $\mat{R}_\alpha$,
$\ver{E}$ and the angular momenta \mat{L}\ and \mat{S}, as opposed to
the corresponding quantities in the gauge $\mat{\xi} = 0$ (the
laboratory frame) which are denoted $\mat{r}_\alpha$, \ver{e},
\mat{l}, \mat{s}. Thus $\S_a(\{\mat{R}_\alpha\}) = 0$ but, in general,
$\S_a(\{\mat{r}_\alpha\}) \neq 0$.  The gauge conditions
(\ref{eq:linear}) select a reference frame rotating so that the linear
combinations of coordinates $\S_a$ vanish for all $t$.  If we choose,
for instance, all coefficients in (\ref{eq:linear}) vanishing except
for $\Gamma_{11Y} = \Gamma_{21Z} = \Gamma_{32Y} = 1$, the coordinate
frame must rotate together with particles 1 and 2 so that 1 is on the
$X$ axis and 2 on the $X-Z$ plane for all $t$.  The formalism in these
linear gauges is entirely analogous to that of non-Abelian Yang-Mills
theories in linear non-covariant gauges, such as the Coulomb or axial
gauges, in which the fields are also constrained by linear relations
\cite{chr}(see also \cite{lnz,grv,tdr}).

For the functions $\S_a(\{\mat{r}_\alpha\})$ to be admissible as gauge
conditions they must not be rotationally invariant.  The variation of
$\S_a$ under an infinitesimal rotation is $\delta\S_a =
\Q_{ak}\delta\theta_k$, with
\begin{equation}
  \label{eq:Qai}
  \Q_{ai}(\{\mat{R}_\gamma\}) = \sum_{\beta=1}^N m_\beta
  \Gamma_{a\beta j} \varepsilon_{jik} R_{\beta k}~, 
  \qquad a=1,2,3.
\end{equation}
The requirement that $\S_a$ must not be invariant under infinitesimal
rotations is therefore satisfied if the matrix $\Q_{ai}$ is not
singular on the gauge manifold.  Thus, the following equations must
be simultaneously satisfied,
\begin{equation}
  \label{eq:simul}
  \S_a(\{\mat{R}_\alpha\}) = 0~, \quad a=1,2,3,
  \qquad
  \det\left(\Q_{bj}\left(\{\mat{R}_\alpha\right)\}\right) \neq 0~,
\end{equation}
except possibly at exceptional configurations at which the gauge is
singular, $\det\Q=0$, such as $\mat{R}_\alpha = 0$ for all $\alpha$.
Furthermore, without loss of generality, we assume that the gauge
coefficients have been orthogonalized so that
\begin{equation}
  \label{eq:ortho}
  \sum_{\alpha=1}^N m_\alpha \Gamma_{a\alpha j} \Gamma_{b\alpha j} =
  \R_a^2 \delta_{ab}~,
  \qquad 
  \R_a^2 \equiv \sum_{\alpha=1}^N m_\alpha \Gamma_{a\alpha j}
  \Gamma_{a\alpha j} > 0~,
  \qquad
  1 \leq a,b \leq 3~.
\end{equation}

The gauge transformation from the gauge $\mat{\xi} = 0$ to the gauge
$\S_a = 0$ is of the form (\ref{eq:trans}),
\begin{equation}
  \label{eq:trans2}
  \mat{R}_\alpha = \mat{U}\mat{r}_\alpha~,
  \qquad
  \ver{E}= \mat{U}\ver{e}~,
  \qquad
  \mat{\xi} = \matd{U} \mat{U}^\dagger~. 
\end{equation}
The orthogonal matrix \mat{U}\ is parametrized by three angles
$\{\theta_a\}_{a=1}^3$.  Although our approach and results do not
depend on any specific parametrization of the rotation group, some
parametrization-dependent quantities, such as the momenta
$p_{\theta_a}$ conjugate to $\theta_a$, are physically meaningful and
play an important r\^ole in some intermediate calculations.  All the
information we will need about the parametrization of \mat{U}\ is
encoded in the matrices \mat{\Lambda}\ and \mat{\lambda}\ defined by 
\begin{equation}
  \label{eq:lambdas}
  \frac{\partial\mat{U}}{\partial \theta_a} \mat{U}^\dagger =
  \Lambda_{ai} \mat{T}_i~,
  \qquad
  \mat{U}^\dagger \frac{\partial\mat{U}}{\partial \theta_a} =
  \lambda_{ai} \mat{T}_i~,
  \qquad
  a=1,2,3,
\end{equation}
where the $\mat{T}_j$ are the standard generators of the $so(3)$
algebra, $(\mat{T}_j)_{ik}=\varepsilon_{ijk}$.  The three matrices
$\partial\mat{U}/\partial\theta_a \mat{U}^\dagger$, $a=1,2,3$, must be
a basis of $so(3)$ for all values of $\{\theta_b\}$, if the
parametrization is to be well defined.  Thus, the matrix
$\Lambda_{ai}$ is invertible and, analogously, so is $\lambda_{ai}$.
{F}rom the unimodularity of \mat{U}\ it follows that $\mat{U}^\dagger 
\mat{T}_i \mat{U} = U_{ij} \mat{T}_j$ and then, from
(\ref{eq:lambdas}), $\lambda_{aj} = \Lambda_{ai} U_{ij}$.  We can
express \mat{\xi}\ in terms of $\theta_a$ and their time derivatives
as
\begin{equation}
  \label{eq:xi}
  \xi_{ik} = \sum_{a=1}^3 \dot{\theta}_a \Lambda_{aj}
  \varepsilon_{ijk}~,
  \quad\text{or}\quad
  \wt{\xi}_j = \sum_{a=1}^3 \dot{\theta}_a \Lambda_{aj}~.
\end{equation}
\mat{\xi}\ in this gauge can also be written in terms of
$\mat{R}_\alpha$, $\ver{E}$, and their time derivatives from the
constraint equation $\mat{L}+\mat{S} = 0$.  The resulting expression,
unlike (\ref{eq:xi}), would be valid only within the constrained
subspace. 

Through (\ref{eq:trans2}), the gauge conditions determine the time
dependence of $\{\theta_a\}$ so that, given a trajectory
$(\{\mat{r}_\alpha(t)\},\ver{e}(t))$ of the system in the gauge
$\mat{\xi} = 0$, we have $\S_a(\{\mat{R}_\alpha(t)\}) =
\S_a(\{\mat{U}(\{\theta_a(t)\})\mat{r}_\alpha(t)\}) = 0$ for all $t$. 
We view (\ref{eq:trans2}) as a coordinate transformation in
configuration space, specifying the new coordinates
$\{\mat{R}_\alpha(\{\mat{r}_\alpha\})\}$,
$\{\theta_a(\{\mat{r}_\alpha\})\}$,
$\ver{E}(\{\mat{r}_\alpha\},\ver{e})$ in terms of the original ones
$\{\mat{r}_\alpha\},\ver{e}$.  The number of independent variables is
the same in both sets, since the $N$ position vectors
$\{\mat{R}_\alpha\}$ are restricted by the three linear conditions
$\S_a(\{\mat{R}_\alpha\}) = 0$.  {F}rom (\ref{eq:trans2}) we then
have, 
\begin{equation}
  \label{eq:dR/dr}
  \frac{\partial R_{\beta i}}{\partial r_{\alpha j}} =
  \delta_{\alpha\beta} U_{ij} + \frac{\partial U_{ik}}{\partial
  r_{\alpha j}} U_{lk} R_{\beta l}~.
\end{equation}
Substituting (\ref{eq:dR/dr}) into the relation $\partial
\S_a(\{\mat{R}_\beta\})/\partial r_{\alpha j} = 0$, and using the
definition (\ref{eq:Qai}) for \Q, the assumption (\ref{eq:simul}) that
it is invertible on the gauge manifold, and the antisymmetry of
$(\partial U_{ik}/\partial r_{\alpha j} U_{lk})$ in $i$ and $l$, we
obtain the relation
\begin{equation}
  \label{eq:Ufund}
  \frac{\partial U_{ik}}{\partial r_{\alpha j}} U_{lk} = \sum_{a=1}^3
  \varepsilon_{ilm} \Q_{ma}^{-1} m_\alpha \Gamma_{a\alpha n} U_{nj}~,
\end{equation}
which expresses $\partial\mat{U}/\partial r_{\alpha j}
\mat{U}^\dagger$ in terms of $\{\mat{R}_\gamma\}$ and $\{\theta_b\}$.
This expression characterizes the dependence of \mat{U} on
$\{\mat{r}_\alpha\}$, and will be important below, especially in the
discussion of angular momentum (see sect.\ \ref{sec:angmom}).  Unlike
the two-dimensional case \cite{bou} in which the explicit form of
\mat{U}\ is easily found, (\ref{eq:Ufund}) does not give us explicit
information about possible Gribov ambiguities of this gauge.  Those
ambiguities are analyzed below (sect.\ \ref{sec:hilbert}), in
connection with the derivation of the Hilbert-space inner product in
this gauge.

The Lagrangian in this gauge is given by \Ll\ in (\ref{eq:glag}) with
$\mat{r}_\alpha$ and \ver{e}\ substituted by $\mat{R}_\alpha$ and
\ver{E}, according to our convention.  Due to the relation
(\ref{eq:xi}) between \mat{\xi}\ and $\dot{\theta}_a$, we can use
$\{\mat{R}_\alpha\}$, $\{\theta_a\}$, \ver{E}\ as dynamical variables,
the Lagrangian in terms of them being obtained by substituting
$\mat{\xi} = \matd{U}\mat{U}^\dagger$ in (\ref{eq:glag}).  Formulating
the theory in those variables, however, would result in momenta
$p_{\theta_a}$ conjugate to $\theta_a$ which are linearly related to
$\mat{J} = \mat{L} + \mat{S}$, not just \mat{L}.  Furthermore, in the
Hamiltonian formulation we have $[L_i, \ver{E}] \neq 0 = [J_i,
\ver{E}]$.  Thus, the gauge transformation (\ref{eq:trans2}) ``mixes''
the particle degrees of freedom $\{\mat{R}_\alpha\}$ and
$\{\theta_a\}$ with the rotator degrees of freedom \ver{E}. We can
avoid such mixing by describing the rigid rotator in terms of its
position versor in the lab frame.  Once the dynamical variables in
(\ref{eq:glag}) have been appropriately capitalized, we set $\mat{\xi}
= \matd{U}\mat{U}^\dagger$ and $\ver{E} = \mat{U}\ver{e}$ to obtain,
\begin{gather}
  \label{eq:glag2}
    \Ll  = \Lln + \Llrt, \qquad \Llrt =\frac{\I}{2} \verd{e}^{\,2}~, 
    \\
    \Lln  = \frac{1}{2} \sum_{\alpha=1}^N m_\alpha \matd{R}_\alpha^2
    + \frac{1}{2} \sum_{\alpha=1}^N m_\alpha \left(\mat{R}_\alpha^2
      \delta_{ij} - R_{\alpha i} R_{\alpha j}\right) \sum_{c,d=1}^3
    \Lambda_{ci} \Lambda_{dj} \dot{\theta}_c \dot{\theta}_d
    - \sum_{\alpha=1}^N m_\alpha \varepsilon_{ijk} R_{\alpha j}
    \dot{R}_{\alpha k} \sum_{c=1}^3 \Lambda_{ci}\dot{\theta}_c - \V~.
    \nonumber
\end{gather}
This Lagrangian, with the gauge conditions (\ref{eq:linear}) holding
as strong (operator) equalities and the constraint $\mat{J} = 0$ valid
as a weak (state space) equality describes the same dynamics as
(\ref{eq:glag}).  The formulation based on (\ref{eq:glag2}), with
$\{\mat{R}_\alpha\}$, $\{\theta_a\}$, \ver{e}\ as dynamical variables,
closely follows the treatment of non-Abelian Yang-Mills theories in
non-covariant linear gauges given in \cite{chr}.

\subsection{Angular and linear momenta}
\label{sec:angmom}

In the quantum theory in the gauge $\mat{\xi} = 0$, as discussed in
section \ref{sec:labo}, the angular momentum operator \mat{l}\ 
satisfies the usual commutator algebra.  Using (\ref{eq:Ufund}), the
definition (\ref{eq:Qai}) of \Q, and the unimodularity of \mat{U}, we
obtain,
\begin{equation}
  \label{eq:commlU}
  [l_i,U_{jk}] = \sum_{\alpha=1}^N \varepsilon_{ilm} r_{\alpha
  l}\frac{1}{i} \frac{\partial U_{jk}}{\partial r_{\alpha m}} = i
  \varepsilon_{ikn} U_{jn}~. 
\end{equation}
{F}rom (\ref{eq:commlU}), using $\mat{L} = \mat{U}\mat{l}$ and 
$\mat{R}_\alpha = \mat{U} \mat{r}_\alpha$, we get,
\begin{equation}\label{eq:commlL}
  \begin{gathered}  
    \left.[L_i,U_{jk}]\right. = -i \varepsilon_{ijn} U_{nk}~,
    \quad
    [l_i, R_{\alpha j}] = 0 = [L_i, R_{\alpha j}]~,\\
    [l_i,l_j] = i \varepsilon_{ijk}l_k~,
    \quad
    [l_i,L_j] = 0~,
    \quad
    [L_i,L_j] = -i \varepsilon_{ijk}L_k~.
  \end{gathered}
\end{equation}
As expected, the particle position vectors $\mat{R}_\alpha$ in this
gauge are rotation invariant.  The commutators among components of
\mat{l}\ and \mat{L}\ are the same as for a rigid body, with \mat{l}\ 
the angular momentum in the laboratory and \mat{L}\ in the body frame.
Furthermore, taking into account the commutators (\ref{eq:xi0}) for
\mat{s}, $[\mat{s}, \mat{U}]=0,$ $\mat{S}=\mat{U}\mat{s}$, and
$\mat{J} \equiv \mat{L} + \mat{S} = \mat{U} \mat{j}$, we have,
\begin{equation}
  \label{eq:commLSJ}
  \begin{gathered}
    \left.[S_i,S_j]\right. = i \varepsilon_{ijk} S_k~,
    \quad
    [L_i,S_j] = -i \varepsilon_{ijk} S_k~,\\
    [J_i,J_j] = -i \varepsilon_{ijk}J_k~,
    \quad
    [J_i,L_j] = -i \varepsilon_{ijk}J_k~,
    \quad
    [J_i,S_j] = 0~.
  \end{gathered}
\end{equation}
Notice that $[\mat{L},\mat{S}] \neq 0$, due to the dependence of
\mat{S}\ on the angles $\{\theta_a\}$.  The classical expressions for
\mat{L}\ and $p_{\theta a}$ follow immediately from the Lagrangian
(\ref{eq:glag2}) and (\ref{eq:angmom}),
\begin{equation}
  \label{eq:pthetaL}
  p_{\theta a} = -\Lambda_{ai} L_i~,
  \qquad
  L_i = \sum_{\alpha=1}^N
  m_\alpha \varepsilon_{ijk} R_{\alpha j} \dot{R}_{\alpha k}
  - \sum_{\alpha=1}^N m_\alpha \left( \mat{R}_\alpha^2
  \delta_{ij} - 
  R_{\alpha i} R_{\alpha j}\right) \wt{\xi}_j ~, 
\end{equation}
with \mat{\wt{\xi}}\ given by (\ref{eq:xi}).  By using the identity
$\sum_{c=1}^3 \partial\Lambda_{ci}/\partial\theta_a \dot{\theta}_c =
\dot{\Lambda}_{ai} + \Lambda_{aj} \varepsilon_{ijk} \sum_{c=1}^3
\dot{\theta}_c \Lambda_{ck},$ which follows from the definition
(\ref{eq:lambdas}) of $\Lambda_{ai}$, the equation of motion for
$\theta_a$ from the Lagrangian (\ref{eq:glag2}) can be reduced to the
form $D_t\mat{L}=0$, in agreement with (\ref{eq:covcon}).  In the
quantum theory,
\begin{equation}
  \label{eq:qpthetaL}
  p_{\theta a} = \frac{1}{i}\frac{\partial}{\partial\theta_a}~,
  \qquad
  L_i = \sum_{a=1}^3 \Lambda_{ia}^{-1} i
  \frac{\partial}{\partial\theta_a}~. 
\end{equation}
Similarly, $p_{\theta a} = -\lambda_{ai} l_i$ and $l_i = \sum_{a=1}^3
\lambda_{ia}^{-1} i \partial/\partial\theta_a.$ Eqs.\ 
(\ref{eq:commlU})-(\ref{eq:qpthetaL}) are completely analogous to the
relations among color currents in Yang-Mills theories in non-covariant
gauges (see eqs.\ (4.44)-(4.48) and (4.55) in \cite{chr}).

In the classical theory we obtain the momenta $\mat{P}_\alpha$
conjugate to $\mat{R}_\alpha$ by differentiating (\ref{eq:glag2}) (or,
equivalently, (\ref{eq:glag})) with respect to $\matd{R}_\alpha$ under
the constraints $\dot{\S}_a(\{\mat{R}_\alpha\}) =
\S_a(\{\matd{R}_\alpha\}) = 0$ to obtain,
\begin{equation}
  \label{eq:mome}
  P_{\alpha i} = m_\alpha \dot{R}_{\alpha i} - m_\alpha \wt{\xi}_j
  \left( \varepsilon_{ijk} R_{\alpha k} - \sum_{b=1}^3
  \frac{1}{\R_b^2} \Q_{bj} \Gamma_{b\alpha i}
  \right)
\end{equation}
with $\Q_{bj}$ and $\R_b^2$ defined in (\ref{eq:Qai}) and
(\ref{eq:ortho}).  These momenta are consistent with the gauge
condition, since they satisfy
\begin{equation}
  \label{eq:pgauge}
  0 = \sum_{\beta=1}^N \Gamma_{a\beta j} P_{\beta j} = \S_a\left(
  \left\{ \mat{P}_\alpha/m_\alpha \right\}\right)~.
\end{equation}
{F}rom the transformation (\ref{eq:trans2}) we can obtain the relation 
between the velocities $\{\matd{r}_\alpha\}$ in the gauge $\mat{\xi} =
0$, and those in the gauge $\S_a = 0$, $\{\matd{R}_\alpha\}$,
$\{\dot{\theta}_a\}$.  Correspondingly, we can express the momenta
$\{\mat{p}_\alpha\}$ in one gauge in terms of the momenta
$\{\mat{P}_\alpha\}$ and \mat{L}\ in the other,
\begin{equation}
  \label{eq:ptrans}
  p_{\alpha j} = U_{kj} \left( P_{\alpha k} + \sum_{a=1}^3 m_\alpha
    \Gamma_{a\alpha k} \Q^{-1}_{na} (L_n - \Lambda_n) \right)~,
  \qquad \text{with} \quad
  \Lambda_n \equiv \sum_{\gamma=1}^N \varepsilon_{npq} R_{\gamma p}
    P_{\gamma q}~.
\end{equation}
The quantity\footnote{There should be no possibility of confusion
  between the 3-component operator $\mat{\Lambda}$ defined in
  (\ref{eq:ptrans}) and the $3\times 3$ matrix $\Lambda_{ai}$ defined
  in (\ref{eq:lambdas}).} 
$\Lambda_n$ defined by this equation has the appearance
of an angular momentum but, as shown below, it does not satisfy the
$so(3)$ commutation relations in general.  With the transformation
(\ref{eq:ptrans}) for momenta we obtain from $\Hn$ in (\ref{eq:xi0})
the classical Hamiltonian for the particle system in this gauge, 
\begin{equation}
  \label{eq:hamS}
  \Hn = \sum_{\alpha=1}^N \frac{1}{2 m_\alpha} \mat{P}_\alpha^2 +
  \frac{1}{2} \sum_{a=1}^3 \R_a^2 \Q^{-1}_{ia} \Q^{-1}_{ja} (L_i -
  \Lambda_i) (L_j - \Lambda_j) + \V~.
\end{equation}
The Hamiltonian $\Hrt$ for the rigid rotator is clearly the same as in
(\ref{eq:xi0}). 

The transformations (\ref{eq:trans2}) and (\ref{eq:ptrans}) can be
inverted, to express $\mat{P}_\alpha$, \mat{L}, $\mat{R}_\alpha$, and
$\theta_a$ in terms of $\mat{p}_\alpha$ and $\mat{r}_\alpha$.  Using
the Poisson brackets (\ref{eq:xi0}), we then get the Poisson brackets
in this gauge.  Alternatively, they can be found as Dirac brackets
\cite{dir} relative to the set of second-class constraints $\S_a(\{
\mat{R}_\alpha\}) = 0 = \S_a(\{ \mat{P}_\alpha/m_\alpha \})$,
$a=1,2,3$.  The results, written in the notation of quantum
commutators, are
\begin{equation}
  \label{eq:comm}
  [R_{\alpha i}, P_{\beta j}] = i \left(\delta_{\alpha\beta}
  \delta_{ij} - \sum_{a=1}^3 \frac{m_\beta}{\R_a^2} \Gamma_{a\alpha i}
  \Gamma_{a\beta j}\right)~.
\end{equation}
All other commutators among $\mat{R}_\alpha$ and $\mat{P}_\beta$
vanish, and $[\mat{L}, \mat{R}_\alpha] = 0 = [\mat{L},
\mat{P}_\beta]$.  {F}rom (\ref{eq:comm}) we get,
\begin{subequations}
\begin{gather}
  {[}\S_a(\{\mat{R}_\alpha\}), P_{\beta j}] = 0 =
  [\S_a(\{\mat{P}_\alpha/m_\alpha\}), R_{\beta j}] = [\Lambda_n,
  \S_a(\{ \mat{R}_\alpha\})]~,\label{eq:comm+a}\\
  [\Lambda_i,\Lambda_j] = i \varepsilon_{ijk} \Lambda_k - i
  \sum_{\alpha=1}^N \sum_{a=1}^3 \frac{1}{\R_a^2} \Gamma_{a\alpha m}
  (\varepsilon_{imn} \Q_{aj} - \varepsilon_{jmn} \Q_{ai}) P_{\alpha
  n}~.  \label{eq:comm+b}
\end{gather}
\end{subequations}
We see that the gauge conditions (\ref{eq:linear}), as well as
(\ref{eq:pgauge}), are operator equations, which can be evaluated
within commutators. We notice also that the definition
(\ref{eq:ptrans}) of \mat{\Lambda}\ is free of ordering problems, even
though the commutators (\ref{eq:comm}) are not canonical, but its
components $\Lambda_i$ in general do not close an angular momentum
algebra, as shown by (\ref{eq:comm+b}).

In the quantum theory, a realization of the commutators
(\ref{eq:comm}) is obtained by defining $\mat{P}_\alpha$ as the
projection of the gradient $\mat{\nabla}_\alpha$ on the gauge
hyperplane $\S_a=0$,
\begin{equation}
  \label{eq:momop}
  P_{\alpha j} = \frac{1}{i} \frac{\partial}{\partial R_{\alpha j}} -
  \sum_{a=1}^3 m_\alpha\Gamma_{a\alpha j} \frac{1}{\R_a^2} 
  \sum_{\beta=1}^N \Gamma_{a\beta k} \frac{1}{i}
  \frac{\partial}{\partial R_{\beta k}}~.
\end{equation}
These operators satisfy both (\ref{eq:comm}) and the gauge condition
(\ref{eq:pgauge}).  They also satisfy relation (\ref{eq:ptrans})
which, with $p_{\alpha j} = -i \partial/\partial r_{\alpha j}$, is
simply the chain rule for derivatives with respect to variables
related by the transformation (\ref{eq:trans2}).  Relations exactly
analogous to (\ref{eq:comm}) and (\ref{eq:momop}) hold in Yang-Mills
theories (compare (\ref{eq:momop}) with the eq.\ between (6.13)
and (6.14) in \cite{chr}).

\subsection{Quantum Hamiltonian}
\label{sec:qham}

The classical Hamiltonian in this gauge, (\ref{eq:hamS}), was obtained
from \Hn\ in the gauge $\mat{\xi} = 0$ by using the transformation
(\ref{eq:ptrans}).  The Hamiltonian operator can in principle be
computed in a similar fashion, essentially by squaring
(\ref{eq:ptrans}) as an operator equation.  That procedure works
satisfactorily in the two-dimensional case \cite{bou}, but in three
dimensions a more systematic approach is needed in order to handle the
much larger amount of algebra required.  The main difference between
the two cases is that in two dimensions the operator $\Lambda$
analogous to \mat{\Lambda}\ in (\ref{eq:ptrans}) commutes with the
Faddeev-Popov determinant \cite{bou}, but that is not the case in
three dimensions. Following \cite{chr} we will first formulate the
theory in terms of an appropriate set of independent generalized
coordinates and their conjugate momenta.  The results obtained in this
intermediate step, which are of interest by themselves, will be
transformed afterwards to the variables $\{\mat{R}_\alpha\}$ and
$\{\theta_a\}$.

The gauge conditions (\ref{eq:linear}) are defined by $9N$ constants
$\Gamma_{a\alpha i}$, $a=1,2,3$, $\alpha=1,\ldots,N$, which
constitute a set of three vectors $\mat{\Gamma}_a$ with $3N$
components $\Gamma_{a\alpha i}$ each, orthogonalized according to
(\ref{eq:ortho}).  We extend the set $\{\mat{\Gamma}_a\}_{a=1}^3$ to
an orthogonal basis $\{\mat{\Gamma}_a\}_{a=1}^{3N}$ of
$\mathbb{R}^{3N}$ by arbitrarily choosing $3(N-1)$ additional vectors
$\{\mat{\Gamma}_b\}_{b=4}^{3N}$ satisfying the orthogonality and
completeness relations,
\begin{equation}
  \label{eq:orthocomp}
  \sum_{\alpha=1}^N m_\alpha \Gamma_{a\alpha j} \Gamma_{b\alpha j} =
  \R_a^2 \delta_{ab}~,
  \quad
  1 \leq a,b \leq 3 N,
  \qquad
  \sum_{a=1}^{3N} \frac{m_\alpha m_\beta}{\R_a^2} \Gamma_{a\alpha i}
  \Gamma_{a\beta j} = m_\alpha \delta_{\alpha \beta} \delta_{ij}~,
\end{equation}
which generalize (\ref{eq:ortho}).  We assume, for simplicity, that
$\R_4^2 = \cdots = \R_{3N}^2 \equiv \R^2$, with $\R^2>0$ an arbitrary
constant.  Furthermore, we choose all $\Gamma_{a\alpha i}$ to have
dimensions of length, so that $\S_a$, $\Q_{ai}$ and $\R_a^2$ all have
the dimensions of an inertia moment.  We define a set of generalized
coordinates $q_a$, $1 \leq a \leq 3N$, in the laboratory gauge
$\mat{\xi} = 0$ by,
\begin{equation}
  \label{eq:q}
  r_{\alpha i}(t) = \sum_{a=1}^{3N} q_a(t) \frac{\Gamma_{a\alpha
      i}}{\R_a}~,
  \qquad
  q_c(t) = \sum_{\alpha=1}^N \frac{m_\alpha}{\R_c} \Gamma_{c\alpha i}
  r_{\alpha i}(t)~,
  \quad
  1 \leq c \leq 3N~.
\end{equation}
Similarly, we introduce $3N-3$ independent generalized coordinates in
the gauge (\ref{eq:linear}) by,
\begin{equation}
  \label{eq:Q}
  R_{\alpha i}(t) = \sum_{a=4}^{3N} Q_a(t) \Gamma_{a\alpha
      i}~,
  \qquad
  Q_c(t) = \sum_{\alpha=1}^N \frac{m_\alpha}{\R^2} \Gamma_{c\alpha i}
  R_{\alpha i}(t)~,
  \quad
  4 \leq c \leq 3N~.
\end{equation}
Due to the orthogonality relations (\ref{eq:orthocomp}), the
expression (\ref{eq:Q}) for $\mat{R}_\alpha$ satisfies the gauge
conditions (\ref{eq:linear}).  The dynamics in this gauge are
completely specified by the $3N$ independent variables
$\{\theta_a\}_{a=1}^3$ and $\{Q_a\}_{a=4}^{3N}$, and their conjugate
momenta.  Notice that the normalization of the coordinates $q_a$ and
$Q_a$ is different.  In (\ref{eq:Q}) the $Q_a$ are chosen to be
dimensionless, for later convenience, whereas in (\ref{eq:q}) the
$q_a$ are defined so that the kinetic energy operator takes the
simplest possible form, that of a Laplacian in Cartesian coordinates.

The Hamiltonian $\Hn$ in the laboratory frame, (\ref{eq:xi0}), is
given in terms of $q_a$ by
\begin{equation*}
  \Hn = -\frac{1}{2} \sum_{a=1}^{3N} \frac{\partial^2}{\partial
  q_a^2} + \V~,
\end{equation*}
from whence the expression for \Hn\ in terms of $\{\theta_a\}$ and
$\{Q_a\}$ follows by means of a coordinate transformation.  The
kinetic energy operator then takes the standard form of a Laplacian in
curvilinear coordinates in configuration space.  As shown in appendix
\ref{sec:appa} the result can be written as
\begin{equation}
  \label{eq:hamS1}
  \begin{aligned}
  \Hn &= -\frac{1}{2 \R^2\J} \sum_{a=4}^{3N} \frac{\partial}{\partial
    Q_a} \J \frac{\partial}{\partial Q_a} \\
  &\quad - \frac{1}{2 |\Lambda|\J}
  \left( \frac{1}{\R^2} \sum_{a=4}^{3N} \frac{\partial}{\partial Q_a} 
    \Q_{ai} + \sum_{b=1}^3 \frac{\partial}{\partial \theta_b}
    \Lambda^{-1}_{ib}\right) \N^{-1}_{ij} |\Lambda| \J \left(
    \frac{1}{\R^2} \sum_{c=4}^{3N} \Q_{cj} \frac{\partial}{\partial
  Q_c} + \sum_{d=1}^3 \Lambda^{-1}_{jd}\frac{\partial}{\partial
  \theta_d} \right)  +\V~,   
  \end{aligned}
\end{equation}
where $|\Lambda| = \det(\Lambda_{ai})$ with $\Lambda_{ai}$ defined in
(\ref{eq:lambdas}) and $\J = \det(\mat{\N})^{1/2}$ with
\begin{equation}
  \label{eq:N}
  \N_{hi} = \sum_{c=1}^3 \frac{1}{\R_c^2} \Q_{ch} \Q_{ci}~,
  \qquad
  \N_{jk}^{-1} = \sum_{d=1}^3 \R_d^2 \Q_{jd}^{-1} \Q_{kd}^{-1}~.
\end{equation}
The quantities $\Q_{ai}$ with $4 \leq a \leq 3N$ appearing in
(\ref{eq:hamS1}) are defined as in eq.\ (\ref{eq:Qai}), of which they
are an extension to $a \geq 4$. The inverse matrix $\Q^{-1}_{ia}$,
however, is defined only for $a\leq 3$.  The expression
(\ref{eq:hamS1}) for \Hn\ depends explicitly on the constants
$\Gamma_{a\alpha i}$ with $a \geq 4$ through $\Q_{ai}$ with $a \geq 4$
and on the parametrization of $\mat{U}(\{\theta_a\})$ through
$\Lambda^{-1}_{kc}$, both of which are largely arbitrary.  Those
dependences will disappear once we recast \Hn\ in terms of
$\mat{P}_\alpha$ and $\mat{L}$.

Using the relation, valid for any matrix depending on a parameter,
\begin{equation}
  \label{eq:lemma8}
  \frac{\partial |\Lambda|}{\partial\theta_a} = |\Lambda| \sum_{b=1}^3
  \Lambda^{-1}_{nb} \frac{\partial \Lambda_{bn}}{\partial\theta_a}~, 
\end{equation}
we obtain (compare (4.48) of \cite{chr})
\begin{equation}
  \label{eq:lemma9}
  \sum_{b=1}^3 \left[p_{\theta_b}, |\Lambda|\Lambda^{-1}_{jb}\right] =
  \sum_{b=1}^3  \frac{1}{i} \frac{\partial
  (|\Lambda|\Lambda^{-1}_{jb})}{\partial\theta_b }  = 0~,
\end{equation}
and hence, using (\ref{eq:qpthetaL}),
\begin{equation}
  \label{eq:auxL}
  \frac{1}{|\Lambda|} \sum_{b=1}^3 \frac{\partial}{\partial\theta_b}
  |\Lambda| \Lambda^{-1}_{jb} = \sum_{b=1}^3 \Lambda^{-1}_{jb}
  \frac{\partial}{\partial\theta_b} = -i L_j~.
\end{equation}
Therefore, we can rewrite \Hn\ in terms of \mat{L}\ as
\begin{equation}
  \label{eq:hamS2}
  \begin{aligned}
  \Hn &= -\frac{1}{2 \R^2\J} \sum_{a=4}^{3N} \frac{\partial}{\partial
    Q_a} \J \frac{\partial}{\partial Q_a} \\
  &\quad - \frac{1}{2 \J} \left( \frac{1}{\R^2} \sum_{a=4}^{3N}
    \frac{\partial}{\partial Q_a} \Q_{ai} -i L_i\right) \N^{-1}_{ij}
  \J \left( \frac{1}{\R^2} \sum_{c=4}^{3N} \Q_{cj}
    \frac{\partial}{\partial Q_c} -i L_j\right) +\V~,
  \end{aligned}
\end{equation}
thus eliminating all explicit dependence of \Hn\ on
$\Lambda^{-1}_{ia}$ and the parametrization of
$\mat{U}(\{\theta_a\})$.  Applying the chain rule, from (\ref{eq:Q})
we have,
\begin{equation}
  \label{eq:chainQ}
  \frac{\partial}{\partial Q_c} = \sum_{\alpha=1}^N \Gamma_{c\alpha i}
  \frac{\partial}{\partial R_{\alpha i}}~,
  \quad
  4 \leq c \leq 3N~.
\end{equation}
The fact that the $\mat{R}_\alpha$ are not independent, being related
by (\ref{eq:linear}), does not affect (\ref{eq:chainQ}) because of the
orthogonality relations (\ref{eq:orthocomp}).  Using the definition
(\ref{eq:momop}) of $\mat{P}_{\alpha}$ and the completeness relation
(\ref{eq:orthocomp}) we get,
\begin{equation}
  \label{eq:PQ}
  P_{\beta j} = \sum_{c=4}^{3N} \frac{m_\beta}{\R^2} \Gamma_{c\beta
  j}\frac{1}{i} 
  \frac{\partial}{\partial Q_c}~,
  \qquad
  \frac{1}{i} \frac{\partial}{\partial Q_d} = \sum_{\beta=1}^N
  \Gamma_{d\beta j} P_{\beta j}~,
  \quad
  4 \leq d \leq 3N~,
\end{equation}
and from (\ref{eq:PQ}) and completeness,
\begin{equation}
  \label{eq:LambdaQ}
  \frac{1}{\R^2}
  \sum_{a=4}^{3N} \frac{1}{i}\frac{\partial}{\partial Q_a} \Q_{ai} =   
  \frac{1}{\R^2}
  \sum_{c=4}^{3N} \Q_{ci} \frac{1}{i}\frac{\partial}{\partial Q_c} = 
  \Lambda_i~,
\end{equation}
with $\Lambda_i$ given by (\ref{eq:ptrans}).  Substituting
(\ref{eq:PQ}) and (\ref{eq:LambdaQ}) into \Hn\ in (\ref{eq:hamS2}) we
finally obtain
\begin{equation}
  \label{eq:hamS3}
  \Hn = \sum_{\alpha=1}^N \frac{1}{2m_\alpha\J} P_{\alpha i} \J
  P_{\alpha i} + \frac{1}{2\J} (L_i-\Lambda_i) \N^{-1}_{ij} \J
  (L_j-\Lambda_j) + \V~,
\end{equation}
in which all dependence on $\{\Gamma_{a\alpha i}\}_{a=4}^{3N}$ has
disappeared.  The total Hamiltonian in this gauge is $\H = \Hn +
\Hrt$, with \Hrt\ from (\ref{eq:xi0}).  Since $\mat{s}$ is a constant
of the motion we can let $\I\rightarrow\infty$ with $\mat{s}$ fixed,
so that \Hrt\ vanishes.  We are then left with an $N$-body system
described by \Hn\ and the constraint $\mat{L}|\mbox{ }\rangle =
-\mat{S}|\mbox{ }\rangle = -\mat{U}(\{\theta_a\}) \mat{s} |\mbox{
}\rangle$ on the state space, with $\matd{s}=[\Hn,\mat{s}]=0$.  We now
turn to this constraint equation.

\subsection{Constraint and physical Hilbert space}
\label{sec:constraint}

The wave function in this gauge $\psi(\{\mat{R}_\alpha\},
\{\theta_a\}, \ver{e})$ is required to satisfy the constraint
$\mat{J}|\psi\rangle = (\mat{L}+\mat{S}) |\psi\rangle = 0$,
originating in the equation of motion (\ref{eq:eqmc}).  Expressing
\mat{L}\ in terms of $p_{\theta a}$ as in (\ref{eq:qpthetaL}), and
using $\mat{S} = \mat{Us}$ and (\ref{eq:lambdas}), we can write the
constraint explicitly as
\begin{equation}
  \label{eq:constraint}
  \left(i\frac{\partial}{\partial\theta_a} + \lambda_{ak} s_k\right) 
  \psi(\{\mat{R}_\alpha\}, \{\theta_a\}, \ver{e}) = 0~,
  \qquad
  a=1,2,3.
\end{equation}
We introduce the unitary operator $\U(\{\theta_a\})$ \cite{chr},
depending on $\{\theta_a\}$ and acting on the Hilbert space of the
rigid rotator, which satisfies the analog in that Hilbert space of
(\ref{eq:lambdas}),
\begin{equation}
  \label{eq:U}
  \frac{\partial \U}{\partial\theta_a} \U^\dagger = i \lambda_{ak}
  s_k~,  \qquad a=1,2,3.
\end{equation}
The matrix elements of $\U(\{\theta_a\})$ in the basis of
eigenfuntions of $\mat{s}^2$, $s_z$, $\langle\ver{e}|s,s_z\rangle =
Y_{ss_z}(\ver{e})$, are the matrices $D^s_{s'_zs_z}(\{\theta_a\})$
(given, e.g., in \cite{gal} in terms of Euler angles).  Defining the
physical wave function $\wh{\psi}(\{\mat{R}_\alpha\},\ver{e})$ 
\begin{equation}
  \label{eq:redwave}
  \psi(\{\mat{R}_\alpha\}, \{\theta_a\}, \ver{e}) = \U(\{\theta_a\})
  \wh{\psi}(\{\mat{R}_\alpha\}, \ver{e})~, 
\end{equation}
we see by direct substitution that (\ref{eq:redwave}) is a solution to
the constraint equation (\ref{eq:constraint}) \cite{chr}.  The wave
functions $\wh{\psi}(\{\mat{R}_\alpha\},\ver{e})$ span the physical
Hilbert space of the system.

Some remarks about the form of the solution (\ref{eq:redwave}) to the
constraint are in order.  By definition $\ver{E} =
\mat{U}(\{\theta_a\}) \ver{e}$, and $\mat{S} = \mat{U}(\{\theta_a\})
\mat{s}$.  Since $[s_i, (\ver{e})_j] = i \varepsilon_{ijk}
(\ver{e})_k$, we have $[S_i, (\ver{E})_j] = i \varepsilon_{ijk}
(\ver{E})_k$ and from (\ref{eq:commlU}), $[L_k, (\ver{E})_l] = -i
\varepsilon_{kln} (\ver{E})_n$.  {F}rom these relations, and taking
into account that \mat{s}\ is a differential operator on the Hilbert
space of the rigid rotator, we obtain $\U(\{\theta_a\}) \chi(\ver{e})
= \chi(\ver{E})$ for any $\chi(\ver{e})$.  Thus, (\ref{eq:redwave})
can be rewritten as $\psi(\{\mat{R}_\alpha\}, \{\theta_a\}, \ver{e}) =
\wh{\psi}(\{\mat{R}_\alpha\}, \ver{E})$.  Had we formulated the theory
in terms of $\{\mat{R}_\alpha\}$, $\{\theta_a\}$ and $\ver{E}$, the
momenta conjugate to $\theta_a$ would have been $p_{\theta a} =
-\Lambda_{ai} J_i$, instead of (\ref{eq:pthetaL}), and the constraint
$\mat{J}\psi = 0$ would have led to $\psi$ not depending on
$\{\theta_a\}$, $\psi(\{\mat{R}_\alpha\}, \{\theta_a\}, \ver{E}) =
\wh{\psi}(\{\mat{R}_\alpha\}, \ver{E})$, the same result as
(\ref{eq:redwave}).

{F}rom (\ref{eq:U}) we obtain $[L_k,\U] = -S_k\U$, which leads to 
$\U^\dagger L_k\U = L_k - \U^\dagger S_k \U$.  Either from this last
equation or from the constraint, on the physical wave functions we
have $\U^\dagger L_k\U\wh{\psi} = -\U^\dagger S_k\U\wh{\psi}=
-s_k\wh{\psi}$.  Therefore, in the physical Hilbert space the
Hamiltonian (\ref{eq:hamS3}) acquires the form,
\begin{equation}
  \label{eq:redham}
  \Hnr \equiv \U^\dagger\Hn\U
  = \sum_{\alpha=1}^N \frac{1}{2m_\alpha\J} P_{\alpha i} \J
  P_{\alpha i} + \frac{1}{2\J} (s_i+\Lambda_i) \N^{-1}_{ij} \J
  (s_j+\Lambda_j) + \V~.
\end{equation}
Similarly, we can write \Hnr\ in terms of the independent coordinates
$\{Q_a\}$ and their conjugate momenta, by just substituting $-\mat{s}$
for \mat{L}\ in (\ref{eq:hamS2}).  Thus, once the constraint has been
solved the angular variables enter the dynamics only through the
dependence of \Hnr\ on \mat{s}, the angular momentum of the rigid
rotator in the lab frame.  Notice that although \mat{s}\ is a constant
of motion in the full Hilbert space of the theory, in general
$[\mat{s},\Hnr]\neq 0$ because $[\mat{s},\U]\neq 0$.  Thus, although
$\mat{s}^2$ can always be diagonalized simultaneously with \Hnr\ 
within the physical subspace, $s_z$ in general cannot be diagonalized.
The physical reason is that within the physical subspace the matrix
elements of \mat{s}\ are equal to those of $-\mat{L}$, which is not
conserved.

\subsection{Inner product in Hilbert space}
\label{sec:hilbert}

In order to find the inner product in the gauge $\S_a=0$ we transform
its expression (\ref{eq:xi0}) in the gauge $\mat{\xi} = 0$ by means of
the Faddeev-Popov technique \cite{fad}.  For that purpose we first
find an appropriate resolution of the identity over the group $SO(3)$,
fixing on the way any Gribov ambiguities \cite{grb,lee} inherent in
the gauge conditions. The invariant integration over $SO(3)$ is given
by (see, e.g., \cite{crt})
\begin{equation}
  \label{eq:haar}
  \int_{SO(3)}\hspace{-10pt} dg f(g) = \int \prod_{a=1}^3
  d\theta_a |\Lambda| f\left(g(\{\theta_a\})\rule{0pt}{9pt}\right)~,   
\end{equation}
where on the l.h.s.\ the integration variable $g$ takes values in
$SO(3)$ and $f:SO(3)\rightarrow\mathbb{C}$.  On the r.h.s.\ of
(\ref{eq:haar}) the integration extends over all of parameter space
and $|\Lambda| \equiv \det(\Lambda_{ai})$ with $\Lambda_{ai}$ defined
in (\ref{eq:lambdas}).  {F}rom (\ref{eq:lemma9}) and (\ref{eq:haar}),
the operators \mat{L}\ defined in (\ref{eq:qpthetaL}) are hermitian,
$\int \prod_{a=1}^3 d\theta_a |\Lambda| \psi^* L_k \phi = (\int
\prod_{a=1}^3 d\theta_a |\Lambda| \phi^* L_k \psi)^*$.

\subsubsection{Resolution of the identity.  Singularities of the
  coordinate frame (Gribov ambiguities)}
\label{sec:gribov}

With the integration measure (\ref{eq:haar}), the resolution of the
identity for the gauge (\ref{eq:simul}) takes the form,
\begin{equation}
  \label{eq:idres}
  1 = \int \prod_{a=1}^3 d\theta_a |\Lambda|
  \prod_{b=1}^3 \delta\left(\frac{1}{\R_b}\S_b\left( \{\mat{U}(\theta)
  \mat{r}_\alpha\}\rule{0pt}{9pt}\right)\right) \J\left(
  \{\mat{U}(\theta) \mat{r}_\alpha\}\rule{0pt}{9pt}\right)
  \Theta\left( \J\left(
      \{\mat{U}(\theta) \mat{r}_\alpha\}\rule{0pt}{9pt}\right) \right)
  \Theta(\F)~,
\end{equation}
where the gauge conditions are conventionally written as $\S_b/\R_b$,
and $\J(\{\mat{r}_\alpha\})$ is defined after (\ref{eq:hamS1}).  To
obtain (\ref{eq:idres}), let $\theta_0=\{\theta_{0a}\}$ be a root to
$\S_b(\{\mat{U}(\theta_0)\mat{r}_\alpha\}) = 0$, $b=1,2,3$, for some
fixed $\{\mat{r}_\alpha\}$.  Then,
\begin{equation}
  \label{eq:variation1}
  \S_b\left( \{\mat{U} (\theta_0 + \delta\theta)
  \mat{r}_\alpha\}\rule{0pt}{9pt}\right) = \sum_{\alpha=1}^N
  m_\alpha \Gamma_{b\alpha j} \delta r_{\alpha j}~,~
  \delta r_{\alpha j} = \sum_{c=1}^3 \delta\theta_c \frac{\partial
    U_{jk}}{\delta \theta_c}(\theta_0) r_{\alpha k}
  =
  \sum_{c=1}^3 \delta\theta_c \Lambda_{cn}\varepsilon_{jnl}
    U_{lm}(\theta_0) r_{\alpha m}
\end{equation}
where for the last equality we used (\ref{eq:lambdas}).  
Using (\ref{eq:Qai}), we rewrite (\ref{eq:variation1}) as 
\begin{equation}
  \label{eq:variation3}
  \frac{1}{\R_b} \S_b\left( \{\mat{U}(\theta_0 + \delta\theta)
  \mat{r}_\alpha\}\rule{0pt}{9pt}\right) = \sum_{c=1}^3 \delta\theta_c
  \frac{1}{\R_b}\Lambda_{cn}
  \Q_{bn}(\{\mat{U}(\theta_0)\mat{r}_\alpha\})~,   
\end{equation}
so that,
\begin{equation}
  \label{eq:variation4}
  \det\left(\frac{\delta(\S_b/\R_b)}{\delta\theta_c}\right)_{\theta_0}
  = |\Lambda| \det\left(\frac{1}{\R_b}\Q_{bn}\left(
      \{\mat{U}(\theta_0)
  \mat{r}_\alpha\}\rule{0pt}{9pt}\right)\right)
  = \frac{1}{\prod_{c=1}^3 \R_c} |\Lambda|
  \det\left(\Q_{bn}\left(\{\mat{U}(\theta_0)\mat{r}_\alpha\}\rule{0pt}{9pt}
  \right)\right)~, 
\end{equation}
and therefore
\begin{equation}
  \label{eq:variation5}
  \prod_{b=1}^3
  \delta\left(\frac{1}{\R_b}
    \S_b\left(\{\mat{U}(\theta_0)\mat{r}_\alpha\}\rule{0pt}{9pt}\right)\right)
  = \sum_{\theta_0} \frac{\prod_{c=1}^3\R_c}{|\Lambda||\det(\Q)|}
  \prod_{a=1}^3 \delta(\theta_a - \theta_{0a})~,
\end{equation}
where the sum extends over all roots $\theta_0$.  In order for the
l.h.s.\ of (\ref{eq:idres}) to be 1, $\theta_0$ must be unique.  The
linear gauge conditions, however, have in general Gribov ambiguities
leading to a discrete set of roots $\theta_0$. We assume that we have
chosen the parametrization $\mat{U}(\{\theta_a\})$ so that $|\Lambda|
> 0$.  But $\det(\mat{\Q})$ can vanish at some configurations
$\{\mat{r}_\alpha\}$ at which the gauge conditions are singular.
Given a configuration $\{\mat{R}^\prime_\alpha =
\mat{U}(\{\theta_a^\prime\})\mat{r}_\alpha, \theta_a^\prime\}$ with
$\det(\mat{\Q})<0$, we can always find a gauge-equivalent one
$\{\mat{R}_\alpha = \mat{U}(\{\theta_a\}) \mat{r}_\alpha, \theta_a\}$
with $\det(\mat{\Q})>0$.  Thus, we restrict ourselves to those
configurations satisfying,
\begin{equation}
  \label{eq:ambigu}
  0<\frac{\det(\mat{\Q})}{\prod_{c=1}^3\R_c} = (\det(\mat{\N}))^{1/2}
  \equiv \J~.
\end{equation}
If this supplementary condition were enough to remove all ambiguities,
together with (\ref{eq:variation5}) it would lead to (\ref{eq:idres}).
Unlike the two-dimensional case \cite{bou}, however, choosing the sign
of the Faddeev-Popov determinant \J\ is in general not enough to
remove the ambiguities.  Further supplementary conditions may be
required which are of the form $F_1>0,\ldots,F_r>0$, where the $F_j$
are $r$ functions of the particle coordinates (as many as necessary to
fix the gauge), such that for each $j$ it is true that $F_j=0$ implies
$\J = 0$.  This set of additional supplementary conditions is
symbolized by the factor $\Theta(\F)$ in (\ref{eq:idres}).

A simple example will illustrate the previous discussion.  Assume that
for a system of $N\geq 3$ particles we want to choose a coordinate
frame rotating so that particle 1 is on the $X$ axis and particle 2 is
on the $X$--$Z$ plane for all $t$.  That frame is not well defined if
particle 1 is at the origin or particle 2 is on the $X$ axis.  The
gauge conditions defining the frame are $\S_1\equiv R_{1Y} =0$,
$\S_2\equiv R_{1Z}=0$, $\S_3\equiv R_{2Y} =0$, leading to
$\det(\mat{\Q}) = -R_{1X}^2 R_{2Z}$.  As expected, $\det(\mat{\Q})=0$
if $R_{1X}=0$ (1 is at the origin) or $R_{2Z}=0$ (2 is on the $X$
axis).  Those singularities stem from the fact that there are four
ways to choose the rotating frame, depending on whether we choose
$R_{1X}\lgt 0$, $R_{2Z}\lgt 0$ for all $t$.  By requiring $\J>0$ we
must have $R_{2Z}<0$, which fixes the ambiguity only partially.  In
order to completely fix the gauge we have to impose a supplementary
condition such as $F_1\equiv R_{1X}>0$.  Clearly, $F_1=0$ implies
$\det(\mat{\Q})=0=\J$.  Alternatively, we may exploit the fact that if
we do not impose the condition $R_{1X}>0$ every configuration is
counted twice (except for those with $R_{1X}=0$ which are counted
once, but they have zero measure and do not contribute to
(\ref{eq:idres})).  Thus, in this example we may omit the factor
$\Theta(\F)$ on the r.h.s.\ of (\ref{eq:idres}) and set the l.h.s.\ to
2.

The general case is analogous to the simple example above.  We either
have to include in (\ref{eq:idres}) the factor $\Theta(\F)$
appropriate to the gauge conditions, or replace it by a factor
$1/(1+N(\{\mat{R}_\alpha\}))$ with $N(\{\mat{R}_\alpha\})$ the number
of gauge-equivalent copies of each configuration $\{\mat{R}_\alpha\}$
satisfying the gauge conditions and $\J>0$ \cite{grb}.  In those cases
in which $N(\{\mat{R}_\alpha\})$ is a constant over all of
configuration space (except maybe for a zero-measure set) we can
omit those factors, absorbing them in the normalization of
the inner product.

\subsubsection{Inner product in Hilbert space}

The inner product in this gauge is straightforward to obtain from
(\ref{eq:xi0}) by using the Faddeev-Popov trick with the resolution of
the identity (\ref{eq:idres}).  We briefly sketch the derivation in
order to highlight the relationship among wave functions in the gauge
$\mat{\xi} = 0$ and those in this gauge.  We rewrite (\ref{eq:xi0})
as, 
\begin{equation}
  \label{eq:ip1}
  \langle\phi | \psi\rangle = \kappa \int \prod_{\beta=1}^N
  d^3\mat{r}_\beta^\prime d^2\ver{e}\; (\phi^*_0
  \psi_0)(\{\mat{U}(\theta_a) \mat{r}_\beta^\prime\},\ver{e})~, 
\end{equation}
where $\kappa > 0$ is a normalization constant to be chosen later (in
(\ref{eq:xi0}), $\kappa=1$), and we performed a change of variables
$\mat{r}^\prime_\beta = \mat{U}(\theta_a)\mat{r}_\beta$ with
$\mat{U}(\theta_a)$ an orthogonal matrix. (In (\ref{eq:ip1}) we
temporarily introduced the notation $\psi_0$ for wave functions in the
gauge $\mat{\xi}=0$ for convenience.)  Inserting (\ref{eq:idres}) in
(\ref{eq:ip1}), exchanging the order of integration, and changing
variables back to $\mat{r}_\beta$, we get
\begin{equation}
  \label{eq:ip2}
  \langle\phi | \psi\rangle = \kappa \int\prod_{a=1}^3 d\theta_a
  |\Lambda| \int \prod_{\beta=1}^N d^3\mat{r}_\beta d^2\ver{e}
  \prod_{b=1}^3 \delta\left(\frac{1}{\R_b}\S_b\right) \J \Theta(\J)
  \Theta(\F) (\phi^*_0 \psi_0)(\{\mat{r}_\beta\},\ver{e})~,  
\end{equation}
where we omitted the argument $\{\mat{r}_\beta\}$ in $\S_b$ and $\J$
for brevity.  We can now choose $\kappa$ to be the reciprocal of the
volume of the rotation group.  Identifying, up to a phase factor, the
physical wave function $\wh{\psi}(\{\mat{R}_\alpha\},\ver{e}) =
\psi_0(\{\mat{R}_\alpha\},\ver{e})$ for $\{\mat{R}_\alpha\}$
satisfying the gauge condition we finally obtain,
\begin{equation}
  \label{eq:ipfinal}
  \langle\phi | \psi\rangle = \int \prod_{\beta=1}^N
  d^3\mat{R}_\beta d^2\ver{e} 
  \prod_{b=1}^3 \delta\left(\frac{1}{\R_b}\S_b\right) \J \Theta(\J)
  \Theta(\F) (\wh{\phi}^* \wh{\psi})(\{\mat{R}_\beta\},\ver{e})~.  
\end{equation}
This gives the inner product in the physical Hilbert space of the
system.  The matrix elements for the Hamiltonian can be computed with
the operator \Hnr\ of (\ref{eq:redham}) and the inner product
(\ref{eq:ipfinal}).  The hermiticity of \Hnr\ with respect to
(\ref{eq:ipfinal}) follows by partial integration, taking into account
that $P_{\alpha i}$ and $\Lambda_i$ are homogeneous first-order
differential operators, with constant coefficients, satisfying
$[P_{\alpha i}, \S_b]=0=[\Lambda_i,\S_b]$, and that $\J\delta(F_j)=0$
since, as mentioned above, $F_j=0$ implies $\J=0$.

We remark also that for states satisfying the constraint
$(\mat{L}+\mat{S})|\psi\rangle = 0$, using the definition
(\ref{eq:redwave}) of the physical wave function and $\U^\dagger
\mat{S}\U = \mat{s}$ we get the equality
\begin{equation}
  \label{eq:ipL}
 \int\prod_{a=1}^3 d\theta_a |\Lambda| \int d\mu\;
 \phi^*(\{\mat{R}_\alpha\}, \{\theta_a\}),\ver{e}) L_i
 \psi(\{\mat{R}_\alpha\}, \{\theta_a\}),\ver{e})  =
 \frac{1}{\kappa} \int d\mu\;
 \wh{\phi}^*(\{\mat{R}_\alpha\},\ver{e}) (-s_i)
 \wh{\psi}(\{\mat{R}_\alpha\},\ver{e})~,
\end{equation}
where we denoted by $d\mu$ the measure appearing in
(\ref{eq:ipfinal}).  The factor $1/\kappa$ appears in (\ref{eq:ipL})
due to the normalization we chose for the inner product in the
physical subspace.  {F}rom (\ref{eq:ipL}) we see that for physical wave
functions the operator $-\mat{s}$ gives the matrix elements of
\mat{L}.

\subsection{Reduced Hamiltonian and Weyl ordering.  Quantum
  potential.}
\label{sec:weylord}

When working in curvilinear coordinates it is often convenient to
redefine the state space by absorbing the Jacobian in the wave
functions, thus eliminating it from the integration measure in the
inner product and from the kinetic energy operator.  That is the case,
for instance, when a perturbative expansion of \J\ contains terms of
many different orders.  Furthermore, the reduced Hamiltonian is easier
to cast into Weyl-ordered form, in which the relation between the
operator and path-integral approaches is straightforward.

The Hamiltonian (\ref{eq:redham}) has been simplified by restricting
it to the physical Hilbert space of gauge-invariant wave functions
$\wh{\psi}$, so it is not of the form of a Laplacian in curvilinear
coordinates.  Thus, we go back to the form (\ref{eq:hamS1}) for $\Hn$,
in which the angles $\{\theta_a\}$ and their conjugate momenta
$p_{\theta a} = -i\partial/\partial \theta_a$ appear explicitly.  {F}rom
(\ref{eq:hamS1}) and (\ref{eq:laplacianWeyl}) we get,
\begin{equation}
  \label{eq:hamW1}
  \begin{aligned}
    \Hnt &\equiv |\Lambda|^\frac{1}{2} \J^\frac{1}{2} \Hn
    |\Lambda|^{-\frac{1}{2}} \J^{-\frac{1}{2}}
    = -\frac{1}{2\R^2} \sum_{a=4}^{3N} \frac{\partial}{\partial Q_a}
    \frac{\partial}{\partial Q_a} - \frac{1}{2\R^4} \sum_{a,b=4}^{3N}
    \left( \Q_{aj} \Q_{bk} \N^{-1}_{jk} \frac{\partial}{\partial Q_a}
      \frac{\partial}{\partial Q_b}\right)_W \\
    &\quad -\frac{1}{2\R^2} \sum_{a=4}^{3N} \sum_{b=1}^3 \left(
      \Q_{aj} \Lambda^{-1}_{kb} \N^{-1}_{jk} \frac{\partial}{\partial
      Q_a} \frac{\partial}{\partial \theta_b} \right)_W
  -\frac{1}{2\R^2} \sum_{a=1}^{3} \sum_{b=4}^{3N} \left(
      \Lambda^{-1}_{ja} \Q_{bk}  \N^{-1}_{jk} \frac{\partial}{\partial
      \theta_a} \frac{\partial}{\partial Q_b} \right)_W \\
 &\quad -\frac{1}{2} \sum_{a,b=1}^{3} \left(
      \Lambda^{-1}_{ja} \Lambda^{-1}_{kb}  \N^{-1}_{jk} \frac{\partial}{\partial
        \theta_a} \frac{\partial}{\partial \theta_b} \right)_W +
    V_Q + \V ~,
  \end{aligned}
\end{equation}
where $V_Q$ is the quantum potential given below (see
(\ref{eq:quanpot})) and $(\cdots)_W$ indicates Weyl-ordering (e.g.,
(\ref{eq:weyldiff})).
Let us introduce the notation
\begin{equation}
  \label{eq:Dnotat}
  \D_{ij}^\alpha \equiv \varepsilon_{ijk} R_{\alpha k} =
  \sum_{a=1}^{3N} \frac{1}{\R_a^2} \Gamma_{a\alpha i} \Q_{aj}~,
\end{equation}
where the second equality follows from (\ref{eq:Qai}) and
(\ref{eq:orthocomp}).  With this definition and relations
(\ref{eq:PQ}) and (\ref{eq:pgauge}), we can rewrite (\ref{eq:hamW1})
in terms of $\mat{P}_\alpha$ operators as,
\begin{equation}
  \label{eq:hamW2}
  \begin{aligned}
    \Hnt &= \sum_{\alpha=1}^N \frac{1}{2 m_\alpha} \mat{P}_\alpha^2 +
    \frac{1}{2} \sum_{\alpha,\beta=1}^N \left( \D_{ij}^\alpha
      \D_{lk}^\beta \N_{jk}^{-1} P_{\alpha i} P_{\beta l} \right)_W
    +\sum_{b=1}^3 \frac{1}{2} \left(\Lambda^{-1}_{kb} \frac{1}{i}
    \frac{\partial}{\partial \theta_b} + \frac{1}{i}
    \frac{\partial}{\partial \theta_b} \Lambda^{-1}_{kb} \right)
    \times\\ 
    &\quad \sum_{\alpha=1}^N \frac{1}{2} \left( \N^{-1}_{jk}
    \D_{ij}^\alpha P_{\alpha i} + P_{\alpha i}  \N^{-1}_{jk}
    \D_{ij}^\alpha \right) +
  \frac{1}{2} \N^{-1}_{jk} \sum_{a,b=1}^3 \left( \Lambda^{-1}_{ja}
    \Lambda^{-1}_{kb} \frac{1}{i} \frac{\partial}{\partial \theta_a}
    \frac{1}{i} \frac{\partial}{\partial \theta_b} \right)_W +
  V_Q + \V ~.
  \end{aligned}
\end{equation}
This operator is to be applied to wave functions of the form
$\wt{\psi}(\{\mat{R_\alpha}\},\{\theta_a\},\ver{e}) = |\Lambda|^{1/2}
\J^{1/2} \psi(\{\mat{R}_\alpha\},\{\theta_a\},\ver{e})$ or, using
(\ref{eq:redwave}),
\begin{equation}
  \label{eq:redwave2}
  \wt{\psi}(\{\mat{R_\alpha}\},\{\theta_a\},\ver{e}) = |\Lambda|^\frac{1}{2}
  \U(\{\theta_a\}) \wth{\psi} (\{\mat{R_\alpha}\},\ver{e})
  \qquad \text{with}\quad
  \wth{\psi} (\{\mat{R_\alpha}\},\ver{e}) \equiv \J^\frac{1}{2}
  \wh{\psi} (\{\mat{R_\alpha}\},\ver{e})~. 
\end{equation}
$\wth{\psi}$ is the reduced form of the physical wave function
$\wh{\psi}$ of (\ref{eq:redwave}).

{F}rom (\ref{eq:lemma9}) we obtain the form of the angular momentum
operator on the first line of (\ref{eq:hamW2}) when applied to wave
functions of the form (\ref{eq:redwave2})
\begin{equation}
  \label{eq:anguangu}
  \sum_{b=1}^3 \frac{1}{2} \left( \Lambda^{-1}_{kb} \frac{1}{i}
  \frac{\partial}{\partial\theta_b} + \frac{1}{i}
  \frac{\partial}{\partial\theta_b} \Lambda^{-1}_{kb} \right)
  \wt{\psi} = -|\Lambda|^\frac{1}{2} L_k \U(\{\theta_a\})
  \wth{\psi} = |\Lambda|^\frac{1}{2} \U(\{\theta_a\}) s_k
  \wth{\psi}~, 
\end{equation}
the second equality following from the discussion immediately above
(\ref{eq:redham}).  After appropriately rearranging the angular
operator on the second line of (\ref{eq:hamW2}) we can apply
(\ref{eq:anguangu}) to it as well and, taking into account that
$\N^{-1}_{ij}$ is symmetric, we get
\begin{equation}
  \label{eq:anguangu2}
  \begin{aligned}
    \frac{1}{2} \N_{jk}^{-1} \sum_{a,b=1}^3 & \left( \Lambda^{-1}_{ja} 
      \Lambda^{-1}_{kb} \frac{1}{i}
      \frac{\partial}{\partial \theta_a} \frac{1}{i}
      \frac{\partial}{\partial \theta_b} \right)_W 
     \wt{\psi} \\
     &= -\frac{1}{8} \N_{jk}^{-1} \sum_{a,b=1}^3 \left\{
       \frac{\partial \Lambda^{-1}_{ja}}{\partial \theta_b}
       \frac{\partial \Lambda^{-1}_{kb}}{\partial \theta_a} + \left(
         \Lambda^{-1}_{ja} \frac{\partial}{\partial\theta_a} +
         \frac{\partial}{\partial\theta_a} \Lambda^{-1}_{ja} \right)
       \left( \Lambda^{-1}_{kb} \frac{\partial}{\partial\theta_b} +
         \frac{\partial}{\partial\theta_b} \Lambda^{-1}_{kb} \right)
     \right\}     \wt{\psi}\\
     &= -\frac{1}{8} |\Lambda|^\frac{1}{2} \U \N_{jk}^{-1}
     \sum_{a,b=1}^3 \frac{\partial \Lambda^{-1}_{ja}}{\partial
     \theta_b} \frac{\partial \Lambda^{-1}_{kb}}{\partial \theta_a}
     \wth{\psi} + \frac{1}{2} |\Lambda|^\frac{1}{2} \U
     \N_{jk}^{-1} s_j s_k \wth{\psi}~.
  \end{aligned}
\end{equation}
Thus, gathering (\ref{eq:hamW2}), (\ref{eq:anguangu}) and
(\ref{eq:anguangu2}) together we obtain
\begin{equation}
  \label{eq:hamW3}
  \begin{aligned}
    \Hnt \wt{\psi} &= |\Lambda|^\frac{1}{2} \U(\{\theta_a\}) \left\{
      \sum_{\alpha=1}^N \frac{1}{2m_\alpha} \mat{P}_\alpha^2 +
      \frac{1}{2} \sum_{\alpha,\beta=1}^N \left( \N^{-1}_{jk}
      \D_{lj}^\alpha \D_{mk}^\beta P_{\alpha l} P_{\beta m}
    \right)_W \right. \\
  &\quad \left. + \frac{1}{2} \sum_{\alpha=1}^N \left( \N^{-1}_{jk} 
      \D_{lj}^\alpha P_{\alpha l} + P_{\alpha l} \N^{-1}_{jk}
      \D_{lj}^\alpha \right) s_k
    + \frac{1}{2} \N^{-1}_{jk} s_j s_k
    - \frac{1}{8} \N^{-1}_{jk} \sum_{a,b=1}^3 \frac{\partial
      \Lambda^{-1}_{ja}}{\partial \theta_b} \frac{\partial
      \Lambda^{-1}_{kb}}{\partial \theta_a} + V_Q + \V \right\}
      \wth{\psi} ~.
  \end{aligned}
\end{equation}
The quantum potential is computed in appendix \ref{sec:quanpot}, with
the result,
\begin{gather}
    V_Q = \frac{1}{8} \N^{-1}_{jk} \sum_{a,b=1}^3 \frac{\partial 
      \Lambda^{-1}_{ja}}{\partial \theta_b} \frac{\partial
      \Lambda^{-1}_{kb}}{\partial \theta_a} + \V_1 + \V_2 ~,
    \qquad
    \V_1 = -\frac{1}{8} \sum_{\alpha=1}^N \sum_{c,d=1}^3 m_\alpha
    \Q^{-1}_{l^\prime c} \Gamma_{c\alpha l} \Q^{-1}_{md}
    \Gamma_{d\alpha k} \varepsilon_{kl^\prime p} \varepsilon_{plm}
      \label{eq:quanpot}\\
    \V_2 = \frac{-1}{8} \sum_{\beta,\gamma=1}^N 
    \varepsilon_{nlk} \N^{-1}_{kh} \varepsilon_{hl^\prime n^\prime}
    \left( \delta_{\beta\gamma} \delta_{l^\prime n} -
    \varepsilon_{l^\prime g s} R_{\gamma s} \sum_{a=1}^3 \Q^{-1}_{ga}
    m_\beta \Gamma_{a\beta n} \right)
  \left( \delta_{\beta\gamma} \delta_{n^\prime l} -
    \varepsilon_{lmp} R_{\beta p} \sum_{b=1}^3 \Q^{-1}_{mb}
    m_\gamma \Gamma_{b\gamma n^\prime} \right).\nonumber
\end{gather}
(Notice that both $\V_{1,2}$ are $\mathcal{O}(|\mat{R}_\alpha|^{-2})$
as $|\mat{R}_\alpha| \rightarrow \infty$).  Thus, the Weyl-ordered,
reduced Hamiltonian $\Hntr$ $= \J^{1/2} \Hnr \J^{-1/2} = \U^\dagger \Hnt
\U$ acting on the space of reduced physical wave functions
$\wth{\psi}(\{\mat{R}_\alpha\},\ver{e})$ is given by,
\begin{equation}
  \label{eq:hamW4}
  \Hntr = \sum_{\alpha=1}^N \frac{1}{2m_\alpha} \mat{P}_\alpha^2 +
  \frac{1}{2} \left(\left(\sum_{\beta=1}^N P_{\beta l} \D_{lj}^\beta +
  s_j\right) \N^{-1}_{jk} \left(\sum_{\gamma=1}^N \D_{mk}^\gamma P_{\gamma m}  +
  s_k\right) \right)_W + \V_1 + \V_2 + \V~,
\end{equation}
where the second term is, explicitly,
\begin{equation}
  \label{eq:secexp}
  \begin{aligned}
    \frac{1}{2} \left(\cdots\rule{0ex}{10pt}\right)_W &= \frac{1}{8}
    \sum_{\beta,\gamma=1}^N \left( \D^\beta_{rj} \N^{-1}_{jk}
      \D^\gamma_{sk} P_{\beta r} P_{\gamma s} +2 P_{\beta r}
      \D^\beta_{rj} \N^{-1}_{jk} \D^\gamma_{sk} P_{\gamma s} +
      P_{\beta r} P_{\gamma s} \D^\beta_{rj} \N^{-1}_{jk}
      \D^\gamma_{sk}
    \right)\\
    &\quad + \frac{1}{4} \sum_{\beta=1}^N \left( \N^{-1}_{jk}
      \D^\beta_{rj} P_{\beta r} + P_{\beta r} \N^{-1}_{jk}
      \D^\beta_{rj} \right) s_k + \frac{1}{2} \N^{-1}_{jk} s_j s_k~.
  \end{aligned}
\end{equation}
Given two states $|\phi\rangle$ and $|\psi\rangle$ represented by the
reduced, physical wave functions $\wth{\phi}$, $\wth{\psi}$,
their inner product is, according to (\ref{eq:ipfinal}) and
(\ref{eq:redwave2}),
\begin{equation}
  \label{eq:ipfinal2}
  \langle\phi | \psi\rangle = \int\prod_{\beta=1}^N d^3\mat{R}_\beta
  d^2\ver{e} \prod_{b=1}^3 \delta\left(\frac{1}{\R_b}\S_b\right)
  \Theta(\J) \Theta(\F) \wth{\phi}^*(\{\mat{R_\beta}\},\ver{e})
  \wth{\psi}(\{\mat{R_\beta}\},\ver{e})~.
\end{equation}
The matrix elements of $\Hntr$ computed with this inner product are,
of course, identical to those of the operator $\Hnr$ of
(\ref{eq:redham}) computed with the inner product (\ref{eq:ipfinal}).   

As pointed out before, there is a close formal analogy between the
results given above for the quantum theory in reference frames defined
by linear conditions and the corresponding results in Yang-Mills
theories in non-covariant linear gauges.  The derivation of the
Hamiltonian given here parallels that of \cite{chr}.  Thus, the
quantum potentials $\V_{1,2}$ from (\ref{eq:quanpot}) are formally
analogous to the corresponding expressions (6.12) and (6.14) in
\cite{chr}.  The kinetic energy operator in (\ref{eq:redham}) and in
the Weyl-ordered form (\ref{eq:hamW4}) are formally equivalent to
(4.62) and (6.15) of \cite{chr}, respectively.  In order to make the
formal analogy clear, we notice that the space derivatives appearing
in field-theoretic expressions must be mapped to zero in the
mechanical case considered here.  Thus, $\Q_{ai}$ from (\ref{eq:Qai})
is the analog of the expression $\Gamma_k \mathfrak{D}_k$ in the
notation of \cite{chr}.  {F}rom (\ref{eq:ptrans}) and (\ref{eq:Dnotat}),
$\Lambda_l = -\sum_{\alpha=1}^N m_\alpha \D^\alpha_{lj} P_{\alpha j}$,
which is the analog of $\mathfrak{D}_i P_i^l \sim -P_i^l
\mathfrak{D}_i$ and, similarly, $\N^{-1}_{ij}$ from (\ref{eq:N}) is
identified with $(\Gamma_k \mathfrak{D}_k)^{-1} (\Gamma_j
\Gamma_j^\dagger) (\mathfrak{D}_k^\dagger \Gamma_k^\dagger)^{-1}$ in
\cite{chr}.

\section{Center of mass motion}
\label{sec:cmm}

In this section and the next one we set $U = 0$ in the Lagrangian and
take into account the translation invariance of (\ref{eq:rlag}) in
order to separate the center-of-mass degrees of freedom.  Since the
center of mass motion is dynamically trivial, we restrict our
treatment to dynamical states with vanishing total momentum.

The Lagrangian (\ref{eq:rlag}) is invariant under time-independent
transformations of the Euclidean group,
\begin{equation}
  \label{eq:eutr}
  \mat{r}^\prime_\alpha = \mat{U}\mat{r}_\alpha + \mat{u}~,
  \qquad
  \ver{e}^\prime = \mat{U} \ver{e}~,
\end{equation}
with \mat{U}\ an orthogonal matrix.  We define the covariant
derivatives
\begin{equation}
  \label{eq:covdev}
D_t \mat{r}_\alpha = \dot{\mat{r}}_\alpha - \mat{\xi}
\mat{r}_\alpha - \mat{\rho}~,
\quad
D_t\ver{e} = \verd{e} - \mat{\xi} \ver{e}~.  
\end{equation}
Under time-dependent transformations $\mat{r}_\alpha$ and
\ver{e}\ transform as in (\ref{eq:eutr}) and,
\begin{equation}
  \label{eq:eutr1}
  \mat{\xi}^\prime = \mat{U}\mat{\xi}\mat{U}^\dagger + \matd{U}
  \mat{U}^\dagger~, 
  \quad
  \mat{\rho}^\prime = \mat{U}\mat{\rho} + \matd{u} - 
   \mat{\xi}^\prime \mat{u}~,
  \quad
  (D_t \mat{r}_\alpha)^\prime = \mat{U} D_t \mat{r}_\alpha~,
  \quad
  (D_t \ver{e})^\prime = \mat{U} D_t \ver{e}~.
\end{equation}
Substituting time derivatives by covariant ones in (\ref{eq:rlag}) we
obtain a Lagrangian invariant under the time-dependent transformations
(\ref{eq:eutr}) and (\ref{eq:eutr1}), 
\begin{equation}
  \label{eq:rhola}
  \Ll = \Lln + \Llrt + \Llcm~,
  \qquad
  \Llcm = \frac{1}{2} \sum_{\alpha=1}^N m_\alpha \mat{\rho}^2 -
  \mat{\rho} \cdot \sum_{\alpha=1}^N m_\alpha \left( \matd{r}_\alpha -
  \mat{\xi} \mat{r}_\alpha\right) ~, 
\end{equation}
where \Lln\ and \Llrt\ have the same form as in (\ref{eq:glag}).  The
equations of motion for $\mat{r}_\alpha$ and \ver{e}\ take the form
(\ref{eq:eqm}) when expressed in terms of covariant derivatives, but
in this case the derivation is slightly more involved because now $D_t
\mat{r}_\alpha$ as given by (\ref{eq:covdev}) does not depend linearly
on $\mat{r}_\alpha$ but, rather, affinely (see appendix
\ref{sec:affine}).  The angular momenta \mat{l}\ and \mat{s}\ are
still given by (\ref{eq:angmom}), with $D_t \mat{r}_\alpha$ from
(\ref{eq:covdev}).  We define $\vlcm = M \vrcm \wedge D_t\vrcm$, where
\vrcm\ is the center-of-mass position, $D_t\vrcm = \vdrcm - \mat{\xi}
\vrcm - \mat{\rho}$, and $M=\sum_{\alpha=1}^N m_\alpha$.  As shown in
appendix \ref{sec:affine}, the equations of motion lead to,
\begin{equation}
  \label{eq:affangmom}
  \left(\frac{d}{dt} - \mat{\xi}\right)(\mat{l} - \vlcm) = 0 = 
  \left(\frac{d}{dt} - \mat{\xi}\right)\mat{s}~,
\end{equation}
which is the same as (\ref{eq:covcon}), with $(\mat{l} - \vlcm)$
instead of \mat{l}.  Thus, the magnitudes of $(\mat{l} - \vlcm)$ and
\mat{s}\ are both conserved and frame-independent.  Due to the fact
that we are now including (time-dependent) translations as symmetries
of the system, the r\^ole played by \mat{l}\ in sections
\ref{sec:model} and \ref{sec:linear} is now played by $(\mat{l} -
\vlcm)$.  The eqs.\ of motion for \mat{\xi}\ and \mat{\rho}\ are now
of the form $\mat{l}+\mat{s}=0$ and $\vpcm=0$ (with $\vpcm = M
D_t\vrcm$).  Thus, in particular $\vlcm=0$ and, like in section
\ref{sec:model}, the total angular momentum of the system vanishes.

If we choose the gauge conditions $\mat{\xi} = 0 = \mat{\rho}$,
corresponding to the laboratory frame, we recover the Lagrangian
(\ref{eq:rlag}), constrained by the eqs.\ of motion for $\mat{\xi}$
and \mat{\rho}\ in this gauge,
\begin{equation}
  \label{eq:oluc}
  \sum_{\alpha=1}^N m_\alpha \mat{r}_\alpha \wedge
  \matd{r}_\alpha + \I \ver{e} \wedge \verd{e} \equiv \mat{l} +
  \mat{s} =0~, 
  \quad
  \sum_{\alpha=1}^N m_\alpha \matd{r}_\alpha =0~.
\end{equation}
These constraints are first class. In the quantum theory they restrict
the state space, $(\mat{l}+\mat{s}) \psi = 0$, $\sum_{\alpha=1}^N
\mat{\nabla}_\alpha \psi = 0$, analogously to Gauss law in Yang-Mills
theories \cite{chr}.  Except for the additional constraint on the
center-of-mass momentum, the quantization in this gauge is carried out
exactly as in section \ref{sec:labo}.

We can now proceed along the same lines as in section
\ref{sec:linear}, imposing on the system the gauge conditions
  \begin{equation}
  \label{eq:gge}
  \S_a(\{\mat{R}_\alpha\}) = 0~,
  \quad
  a=1,2,3,
  \qquad
  \mat{\C}(\{\mat{R}_\alpha\}) \equiv \frac{1}{M}\sum_{\beta=1}^N
  m_\beta \mat{R}_\beta = 0~,
\end{equation}
with $\S_a$ defined in (\ref{eq:linear}).  (\ref{eq:gge}) defines a
reference frame in a particular state of rotation, with origin at the
center of mass.  Like in section \ref{sec:linear}, in the rest of this
section we denote vectors referred to this frame by capital letters,
while lower-case symbols denote lab frame quantities.  The gauge
conditions (\ref{eq:gge}) are not mutually consistent unless $\S_a$
are translation invariant,
\begin{equation}
  \label{eq:trinv}
  \sum_{\alpha=1}^N m_\alpha \Gamma_{a\alpha j} = 0,
  \quad
  a=1,2,3~,
\end{equation}
the condition $\mat{\C}=0$ being clearly rotationally invariant.
Furthermore, we assume that $\S_a$ satisfy (\ref{eq:simul}) and
(\ref{eq:ortho}).
The gauge transformation from the gauge $\mat{\xi} = 0 = \mat{\rho}$ 
is of the form (\ref{eq:eutr})-(\ref{eq:eutr1}), with parameter
$\mat{u} = -\mat{U}\vrcm$, where $\vrcm = \sum_{\alpha=1}^N
m_\alpha/M \mat{r}_{\alpha}$ is the center of mass in the lab
frame,  
\begin{equation}
  \label{eq:eutr2}
  \mat{R}_\alpha = \mat{U} \left(\mat{r}_\alpha - \vrcm\right)~,
  \quad
  \ver{E} = \mat{U} \ver{e}~,
  \quad
  \mat{\xi} = \matd{U} \mat{U}^\dagger~,
  \quad
  \mat{\rho} = -\mat{U}\vdrcm~.
\end{equation}
The transformation (\ref{eq:eutr2}) mixes the particle degrees of
freedom $\mat{R}_\alpha$ with those of the center of mass and the
rigid rotator, just like (\ref{eq:trans2}) did in the
non-translation-invariant case.  In particular, in these variables
$p_{\theta_a}$ is not linearly related to \mat{L}.  That mixing is
avoided, as in section \ref{sec:linear}, by trading the dynamical
variables $\{\mat{R}_\alpha\}$, \ver{E}, \mat{\xi}, \mat{\rho}\ for
$\{\mat{R}_\alpha\}$, \ver{e}, $\{\theta_a\}$, \vrcm.  Substituting
the last three of (\ref{eq:eutr2}) into the Lagrangian
(\ref{eq:rhola}) we get,
\begin{equation}
  \label{eq:rhola2}
  \Ll = \Lln + \Llrt + \Llcm~,
  \qquad
  \Llcm = \frac{1}{2} M \vdrcm^2~,
\end{equation}
with \Lln\ and \Llrt\ now given by (\ref{eq:glag2}). The Lagrangian
(\ref{eq:rhola2}) is supplemented by the gauge conditions
(\ref{eq:gge}) holding as strong (operator) equations, and the
constraints $\mat{J}=\mat{L}+\mat{S}=0$ and $\vpcm \equiv M \vdrcm =
0$ valid as weak (state space) equalities.

We keep the definitions (\ref{eq:lambdas}) of $\Lambda_{ai}$ and
$\lambda_{ai}$ from section \ref{sec:linear}.  The expression
(\ref{eq:xi}) of \mat{\xi}\ in terms of $\{\dot{\theta}_a\}$ then
holds unchanged since $\mat{\xi}=\matd{U}\mat{U}^\dagger$ just like in
section \ref{sec:linear}.  {F}rom \Ll\ in (\ref{eq:rhola2}) we can then
derive the relations (\ref{eq:pthetaL}) for $p_{\theta_a}$ and the
angular momentum \mat{L}\ in this gauge.  The classical expression
(\ref{eq:pthetaL}) for \mat{L}, in turn, together with the
transformation law (\ref{eq:eutr2}) lead to the relation
\begin{equation}
  \label{eq:Ltrans}
  \mat{L} = \mat{U} (\mat{l} - \vlcm)~,
\end{equation}
where the center-of-mass angular momentum in the lab frame is defined
as $\vlcm = \vrcm \wedge \vpcm = M \vrcm \wedge \vdrcm$.  {F}rom the
relation (\ref{eq:eutr2}) between $\mat{R}_\alpha$ and
$\mat{r}_\alpha$ we can derive an expression for $\partial
\mat{U}/\partial r_{\alpha j} \mat{U}^\dagger$ by following the same
steps leading to (\ref{eq:Ufund}) in section \ref{sec:linear}.  The
result is that (\ref{eq:Ufund}) remains valid without modifications
and that, due to the translation invariance condition (\ref{eq:trinv})
for $\S_a$, \mat{U}\ does not depend on \vrcm, 
\begin{equation}
  \label{eq:Ufund2}
  \sum_{\alpha=1}^N \frac{\partial \mat{U}}{\partial \mat{r}_\alpha} =
  0~. 
\end{equation}
In particular, $[\vlcm,\mat{U}]=0$.  {F}rom (\ref{eq:Ufund}), in
turn, the commutator (\ref{eq:commlU}) of \mat{l}\ with \mat{U}\ 
follows.  Thus, with the commutator (\ref{eq:commlU}), the
transformation relations (\ref{eq:eutr2}) and (\ref{eq:Ltrans}), and
(\ref{eq:Ufund2}), we recover all of the commutators (\ref{eq:commlL})
and also,
\begin{equation}
  \label{eq:commlcm}
  [{\lcm}_i, {\lcm}_j] = i \varepsilon_{ijk} {\lcm}_k~,
  \quad
  [l_i, {\lcm}_j] = i \varepsilon_{ijk} {\lcm}_k~,
  \quad
  [{\lcm}_i, U_{jk}] = 0~,
  \quad
  [{\lcm}_i, L_j] = 0~.
\end{equation}
Furthermore, $[{\lcm}_i, s_j] = 0$.  Thus, since $\mat{J}\equiv
\mat{L} + \mat{S} = \mat{U} (\mat{l} -\vlcm+\mat{s})$, with
(\ref{eq:commlcm}) we find $[{\lcm}_i, S_j] = 0 = [{\lcm}_i, J_j]$ and
then all of the commutators (\ref{eq:commLSJ}) follow.  In summary,
with the exception of eq.\ (\ref{eq:Ltrans}), all of the results of
section \ref{sec:angmom} remain valid in this case. 

The relation among linear momenta in the gauge
$\mat{\xi} = 0 = \mat{\rho}$ and the gauge $\S_a = 0 =\mat{\C}$
analogous to (\ref{eq:ptrans}) takes the form,
\begin{equation}
  \label{eq:ptrans2}
    p_{\alpha j} = U_{kj} \left( P_{\alpha k} + \sum_{a=1}^3 m_\alpha
    \Gamma_{a\alpha k} \Q^{-1}_{na} (L_n - \Lambda_n)\right) +
    \frac{m_\alpha}{M} {\pcm}_j 
\end{equation}
with $\mat{\Lambda}$ defined as in (\ref{eq:ptrans}).
Correspondingly, the classical Hamiltonian is given by (\ref{eq:hamS})
with the addition of the center-of-mass kinetic energy $\vpcm^2/(2M)$.
Due to the additional gauge conditions $\mat{\C}=0$ the fundamental
commutators (\ref{eq:comm}) become,
\begin{equation}
  \label{eq:commT}
    [R_{\alpha i}, P_{\beta j}] = i \left(\delta_{\alpha\beta}
  \delta_{ij} - \frac{m_\beta}{M} \delta_{ij} - \sum_{a=1}^3
  \frac{m_\beta}{\R_a^2} \Gamma_{a\alpha i} 
  \Gamma_{a\beta j}\right)~,
\end{equation}
and $[P_{\alpha i}, \S_a(\{\mat{R_\beta}\})] = 0 = [P_{\alpha i},
\mat{\C}(\{\mat{R_\beta}\})]$.  The differential operators realizing
this algebra are obtained in the same way as those in
(\ref{eq:momop}), which is now modified to
\begin{equation}
  \label{eq:momopT}
  P_{\alpha i} = \frac{1}{i} \frac{\partial}{\partial R_{\alpha i}} -
  \sum_{a=1}^3 \frac{m_\alpha}{\R_a^2} \Gamma_{a\alpha i}
  \sum_{\beta=1}^N \Gamma_{a\beta j} \frac{1}{i}
  \frac{\partial}{\partial R_{\beta j}} - \frac{m_\alpha}{M}
  \sum_{\beta=1}^N \frac{1}{i} \frac{\partial}{\partial R_{\beta i}}.
\end{equation}
These operators satisfy $\S_a(\{\mat{P_\alpha}/m_\alpha\}) = 0 =
\sum_{\alpha=1}^N \mat{P}_\alpha$.  They also satisfy
(\ref{eq:ptrans2}), with $p_{\alpha j} = 1/i \partial/\partial
r_{\alpha j}$.  The additional term in (\ref{eq:momopT}) with respect
to (\ref{eq:momop}) does not modify the form of \mat{\Lambda}\ as a
differential operator.  Clearly, the commutators of $\mat{R}_\alpha$ and
$\mat{P}_\alpha$ with \vrcm\ and $\vpcm = 1/i \partial/\partial \vrcm$
vanish strongly.

\subsection{Quantum Hamiltonian}
\label{sec:qham2}

The Hamiltonian operator is obtained by the same procedure as in
section \ref{sec:qham}, with obvious modifications.  We 
omit all calculational details and quote the results only, after
establishing the appropriate notation.

Like in section \ref{sec:qham}, we extend the gauge coefficients
$\Gamma_{a\alpha i}$, $a=1,2,3$, $\alpha=1,\ldots,N$, to an orthogonal
basis of $\mathbb{R}^{3N}$.  That is, we consider an extended set of
coefficients $\Gamma_{a\alpha i}$ with $1\leq a \leq 3N$,
$\alpha=1,\ldots,N$, $i=1,\ldots,3$, satisfying the orthogonality and
completeness relations (\ref{eq:orthocomp}).  For simplicity, like in
section \ref{sec:qham}, we set the normalization constants in
(\ref{eq:orthocomp}) to be independent of $a$ for $4 \leq a \leq 3N$,
$\R_a^2 = \R^2 > 0$ with $\R^2$ a constant at our disposal.
Furthermore, for $3N-2 \leq a \leq 3N$ we choose the coefficients
$\Gamma_{a\alpha j}$ to be independent of $\alpha$.  Specifically, we
set,
\begin{equation}
  \label{eq:gammahigh}
  \Gamma_{a\alpha i} = \frac{\R}{\sqrt{M}} \delta_{(a-3N+3)i}~,
  \quad
  3N-2 \leq a \leq 3N~,
  \quad
  1 \leq \alpha \leq N~.
\end{equation}
With this choice the orthogonality relations (\ref{eq:orthocomp}) with
$3N-2\leq a \leq 3N$ and $1\leq b \leq 3N-3$ read,
\begin{equation}
  \label{eq:orthocomp2}
  \sum_{\alpha=1}^N m_\alpha \Gamma_{b\alpha j} = 0~,
  \quad
  1 \leq b \leq 3N-3~.
\end{equation}
In particular, (\ref{eq:orthocomp2}) contains the translation
invariance conditions (\ref{eq:trinv}) for $\S_b$.   

We can now define the generalized coordinates $\{q_a\}_{a=1}^{3N}$ by
(\ref{eq:q}).  We see that for $3N-2 \leq a \leq 3N$, $q_a = \sqrt{M}
{\rcm}_{(a-3N+3)}$, i.e., up to a normalization constant the last
three $q_a$ are the components of \vrcm.  The analog of (\ref{eq:Q})
is now,
\begin{equation}
  \label{eq:Q2}
  R_{\alpha i}(t) = \sum_{a=4}^{3N-3} Q_a(t) \Gamma_{a\alpha
      i}~,
  \qquad
  Q_c(t) = \sum_{\alpha=1}^N \frac{m_\alpha}{\R^2} \Gamma_{c\alpha i}
  R_{\alpha i}(t),
  \quad
  4 \leq c \leq 3N-3~.
\end{equation}
{F}rom (\ref{eq:orthocomp}) and (\ref{eq:gammahigh}) the expression
(\ref{eq:Q2}) for $\mat{R}_\alpha$ satisfies the gauge conditions
(\ref{eq:gge}).  The dynamics in this gauge is then completely
specified by the $3N$ independent variables $\{\theta_a\}_{a=1}^3$,
$\{Q_a\}_{a=4}^{3N-3}$ and \vrcm, and their conjugate momenta.  In
those variables \Hn\ is given by (\ref{eq:hamS1}) with only two
modifications: first, the sums over indices running up to $3N$ now run
only up to $3N-3$, and second, the addition of the term $-1/(2M)
\partial^2/\partial {\rcm}_j^2$.  The definitions (\ref{eq:N}) of
$\N_{ij}$ and its inverse, and of \J\ and $|\Lambda|$ remain
unchanged.  Similarly, \Hn\ is expressed in terms of $\mat{R}_\alpha$,
their conjugate momenta $\mat{P}_\alpha$, and \mat{L}\ by
(\ref{eq:hamS3}), but now with the momentum operators $\mat{P}_\alpha$
from (\ref{eq:momopT}), and with the addition of the center-of-mass
kinetic energy term.

\subsection{Physical Hilbert space. Inner product. Weyl-ordered
  Hamiltonian} 
\label{sec:constraint2}

The wave function in this gauge
$\psi(\{\mat{R}_\alpha\},\{\theta_a\},\ver{e},\vrcm)$ is required to
satisfy the constraints $\vpcm \psi=0$ and $(\mat{L} + \mat{S})\psi=0$
originating in the equations of motion for \mat{\rho}\ and \mat{\xi}\
from the Lagrangian (\ref{eq:rhola}).  The first constraint is trivial
to solve.  Considering wave functions $\psi$ independent of \vrcm, we
are left with the constraint on the angular variables which, since
$[\vpcm, \mat{L}]=0=[\vpcm,\mat{S}]$, can now be treated exactly as in
section \ref{sec:constraint}.  Using the same notation as in
(\ref{eq:redwave}), the solution to the constraint equations is of
the form,
\begin{equation}
  \label{eq:constraint2}
  \psi(\{\mat{R}_\alpha\},\{\theta_a\},\ver{e},\vrcm) = \U(\{\theta_a\})
  \wh{\psi}(\{\mat{R}_\alpha\}, \ver{e})~. 
\end{equation}
Within the subspace of physical wave functions
$\wh{\psi}(\{\mat{R}_\alpha\}, \ver{e})$ the Hamiltonian $\Hnr \equiv
\U^\dagger\Hn\U$ is given by (\ref{eq:redham}), with the momentum
operators $\mat{P}_\alpha$ from (\ref{eq:momopT}).

The discussion of the inner product from section \ref{sec:hilbert}
requires only minor changes in order to adapt it to the
translation-invariant case.  Besides the resolution of the identity
(\ref{eq:idres}) for the rotational gauge conditions $\S_a=0$, we have
to fix the translational gauge by means of a resolution of the form,
\begin{equation}
  \label{eq:idres2}
  1 = \int d^3u \prod_{i=1}^3 \delta\left( \C_i(\{\mat{r}_\alpha\}) +
  u_i \right)~.  
\end{equation}
Inserting this factor of one together with (\ref{eq:idres}) into the
canonical inner product (\ref{eq:xi0}), we obtain $\langle\phi |
\psi\rangle$ in terms of wave functions in this gauge.  A technical
detail is that, after applying the Faddeev-Popov procedure, the volume
of the symmetry group appears as a prefactor in $\langle\phi |
\psi\rangle$ (see (\ref{eq:ip2})).  In this case, the volume of the
translation group is infinite so an appropriate limiting or
regularization procedure must be applied.  Assuming that has been
done, the resulting inner product in terms of physical wave functions
analogous to (\ref{eq:ipfinal}) is
\begin{equation}
  \label{eq:ipfinal3}
  \langle\phi | \psi\rangle = \int \prod_{\alpha=1}^N
  d^3\mat{R}_\alpha d^2\ver{e} 
  \prod_{a=1}^3 \delta\left(\frac{1}{\R_a}\S_a\right)
  \delta^{(3)}\left(\mat{\C}\right)
  \Theta(\F) \Theta(\J) \J
  (\wh{\phi}^* \wh{\psi})(\{\mat{R}_\alpha\},\ver{e})~.  
\end{equation}
The factor $\Theta(\F)$ is exactly as discussed in section
\ref{sec:gribov}, since the center-of-mass condition does not
introduce further Gribov ambiguities.

We consider, finally, the form of the Weyl-ordered reduced
Hamiltonian.  Taking proper account of translation invariance as
described above, the analysis of section \ref{sec:weylord} remains
valid \emph{mutatis-mutandis.}  Defining the reduced physical wave
functions
\begin{equation}
  \label{eq:redphys}
  \wth{\psi} (\{\mat{R_\alpha}\},\ver{e}) \equiv \J^\frac{1}{2}
  \wh{\psi} (\{\mat{R_\alpha}\},\ver{e})~,   
\end{equation}
with $\wh{\psi}$ as defined in (\ref{eq:constraint2}), the reduced
Hamiltonian $\Hntr = \J^{1/2} \Hnr \J^{-1/2} = \J^{1/2}\U^\dagger \Hnt
\U\J^{-1/2}$ is given in Weyl-ordered form by (\ref{eq:hamW4}), with
$\mat{P}_\alpha$ given by (\ref{eq:momopT}).  The quantum potentials
$\V_{1,2}$, in particular, are still defined as in (\ref{eq:quanpot}).
The reduced inner product is immediately obtained from
(\ref{eq:ipfinal3}) and (\ref{eq:redphys}).

\section{Quasi-rigid systems in the Eckart frame}
\label{sec:quasi}

We assume now that the potential energy \V\ (with $U =0$) has a
minimum for some configuration $\{\mat{z}_\alpha\}$ of the system,
such that $\V_0 \equiv \V(\{\mat{z}_\alpha\}) \leq
\V(\{\mat{r}_\gamma\})$ for all configurations $\{\mat{r}_\gamma\}$.
Due to the invariance of \V\ under the Euclidean group $E_3$ any
configuration $\{\mat{z}_\alpha^\prime\}$ related to
$\{\mat{z}_\alpha\}$ by a transformation of the form (\ref{eq:eutr})
is also a minimum.  Denoting by $\mathcal{M}_\V$ the manifold of
configuration space defined by $\V(\{\mat{r}_\gamma\}) = \V_0$, we
assume that the quotient $\mathcal{M}_\V/E_2$ is a discrete set.  The
configurations of minimal potential energy are therefore rigid.  In
this section we discuss the quantization of the small oscillations of
the system about these rigid equilibrium configurations.  We will
denote by $\{\mat{Z}_\alpha\}$ the unique (up to discrete degeneracy)
minimum of \V\ satisfying,
\begin{equation}
  \label{eq:mini}
  \sum_{\alpha=1}^N m_\alpha Z_{\alpha i} Z_{\alpha j} = 0~,
  \quad i\neq j~,
  \qquad
  \sum_{\alpha=1}^N m_\alpha \mat{Z}_{\alpha} = 0~.
\end{equation}
We will restrict ourselves to considering only systems for which the
inertia tensor for the equilibrium configuration $\{\mat{Z}_\alpha\}$
is non-singular.  The small oscillations of the system are described
by trajectories of the form,
\begin{equation}
  \label{eq:traj}
  \mat{r}_\alpha (t) = \mat{z}_\alpha (t) + \mat{\delta r}_\alpha (t)
  \quad
  \text{with}
  \quad
  \mat{z}_\alpha (t) = \mat{U}(t) \mat{Z}_\alpha + \mat{u}
\end{equation}
for some orthogonal matrix $\mat{U}(t)$ and $\mat{u}$ appropriately
chosen so that $\mat{\delta r}_\alpha (t)$ are small with respect to
their characteristic scale for all $t$. Since we restrict ourselves to
states with vanishing total momentum, the translation vector $\mat{u}$
in (\ref{eq:traj}) must be time-independent.

It is convenient to apply the inverse of the gauge transformation
defined by the second equation in (\ref{eq:traj}) in order to switch
to a reference frame, the ``body frame'' of the rigid equilibrium
configuration, so that
\begin{equation}
  \label{eq:body}
  \mat{r}_\alpha (t) = \mat{Z}_\alpha + \mat{\delta r}_\alpha (t)~.
\end{equation}
This fixes the gauge only to leading order in $\mat{\delta r}_\alpha$.
We fix the residual gauge freedom by imposing a gauge condition on
$\mat{\delta r}_\alpha$, which amounts to correcting the definition
(\ref{eq:mini})-(\ref{eq:body}) of the reference frame by small
quantities of first order.  We choose the origin of the reference
frame at the center of mass, so to first order in $\mat{\delta
  r}_\alpha$ the gauge conditions must be of the form (\ref{eq:gge}).
The choice of the coefficients $\Gamma_{a\alpha i}$ is arbitrary as
long as (\ref{eq:trinv}) is satisfied.  We then have,
\begin{equation}
  \label{eq:ppale}
  \mat{R}_\alpha (t) = \mat{Z}_\alpha + \mat{\delta R}_\alpha (t)~,
  \quad
  \S_a(\{\mat{\delta R}_\alpha\}) = 0~,
  \quad
  \mat{\C}(\{\mat{\delta R}_\alpha\}) = 0~.
\end{equation}
The Eckart frame corresponds to choosing $\Gamma_{a\alpha
  i}=\varepsilon_{aji} Z_{\alpha j}$, $a=1,2,3$ \cite{ek2,gai}.  With
this choice the normalization constants $\R_a^2$, $a=1,2,3$, of
(\ref{eq:ortho}) (and also (\ref{eq:orthocomp}) and
(\ref{eq:orthocomp2})) are given by the inertia moments of the
equilibrium configuration, and the matrix $\Q(\{\mat{R}_{\alpha i}\})$
of (\ref{eq:Qai}) is,
\begin{equation}
  \label{eq:QaiE}
  \Q_{ai}(\{\mat{\delta R}_{\alpha}\}) = \R_a^2 \delta_{ai} + \delta 
  \Q_{ai}(\{\mat{\delta R}_\alpha\})~,
  \quad
  \delta \Q_{ai}(\{\mat{\delta R}_\alpha\}) =
  \sum_{\gamma=1}^N 
  m_\gamma (\mat{Z}_\gamma\cdot\mat{\delta R}_\gamma \delta_{ai} -
  \delta R_{\gamma a} Z_{\gamma i} )~,
  \quad
  a=1,2,3~.
\end{equation}
Thus, since $\R_a^2 \neq 0$ by assumption, for small $\mat{\delta
  R}_\alpha$ the condition $\det(\Q_{ai}(\{\mat{R}_\alpha\}))\neq 0$
is satisfied.

The momentum operators $\mat{P}_\alpha$ and their fundamental
commutators are as given in (\ref{eq:commT}) and (\ref{eq:momopT}),
with the replacement of $\partial/\partial R_{\alpha i}$ by
$\partial/\partial \delta R_{\alpha i}$.  The operator \mat{\Lambda}\
defined in (\ref{eq:ptrans}) can be rewritten in the form, 
\begin{equation}
  \label{eq:residual}
  \Lambda_i = \sum_{\alpha=1}^N \varepsilon_{ijk} \delta R_{\alpha j}
  \frac{1}{i} \frac{\partial}{\partial \delta R_{\alpha k}} -
  \sum_{a=1}^3 \frac{\delta\Q_{ai}}{\R_a^2} \sum_{\alpha=1}^N
  \varepsilon_{ajk} Z_{\alpha j} \frac{1}{i}
  \frac{\partial}{\partial \delta R_{\alpha k}}~.
\end{equation}
As expected in this gauge \cite{ek2,bou}, its coefficients are of
$\mathcal{O}(\mat{\delta R}_\alpha)$.  Similarly, the operators
$\sum_{\alpha=1}^N \D_{ij}^\alpha P_{\alpha i}$ and $\sum_{\alpha=1}^N
P_{\alpha i} \D_{ij}^\alpha$ appearing in $\Hntr$ are of
$\mathcal{O}(\mat{\delta R}_\alpha)$.  The Hamiltonian \Hntr, given by
(\ref{eq:hamW4}) with $\mat{P}_\alpha$ from (\ref{eq:momopT}), is
obtained perturbatively by expanding (\ref{eq:hamW4}) in powers of
$\mat{\delta R}_\alpha$.  {F}rom this point of view, the elimination of
the Jacobian \J\ from the kinetic energy, as indicated in sections
\ref{sec:weylord} and \ref{sec:constraint2}, is particularly
convenient in perturbation theory.  The inner product, finally, is
given by (\ref{eq:ipfinal3}) with the modification (\ref{eq:redphys}),
and with $\mat{\delta R}_\alpha$ as integration variable.  Once the
equilibrium configuration $\{\mat{Z}_\alpha\}$ has been chosen, its
body frame is uniquely fixed.  The Eckart frame is then equally well
defined as long as $\{\mat{\delta R}_\alpha\}$ are small.  Thus,
except in those cases in which the equilibrium configuration
$\{\mat{Z}_\alpha\}$ is exceptionally close to a zero of \J, we can
neglect the factor $\Theta(\J)$ in (\ref{eq:ipfinal3}) since large
displacements $\mat{\delta R}_\alpha$ which could drive \J\ to zero
should be exponentially suppressed by the wave function.  Similarly,
we also expect to be able to neglect $\Theta(\F)$ in
(\ref{eq:ipfinal3}) in perturbation theory.

As a minimal illustration and consistency check of the formalism we
analyze below a simple example with $N=3$, and briefly comment on the
$N=4$ case.  The natural variables for quasi-rigid systems are normal
coordinates, so below we recast the Hamiltonian in terms of those
coordinates.  We remark, however, that the results of the previous
sections are more general than the simple examples considered here,
and can be applied to non-quasi-rigid $N$-body sytems, both in the
operator and path integral formalims.

\subsection{A simple example with \mat{N=3}}
\label{sec:simple3}

The simplest possible model, within our assumptions, consists of three
particles of equal mass $m$ interacting through a two-body potential
\V\ as in (\ref{eq:rlag}), with $U=0$ and $V_{\alpha\beta}=V$
independent of $\alpha$, $\beta$.  $V(r)$ is assumed to have an
absolute minimum at $r=a>0$.  The classical equilibrium configurations
are then those in which the particles lie at relative rest on the
vertices of an equilateral triangle of side $a$.  An equilibrium
configuration satisfying (\ref{eq:mini}), unique up to permutations of
the particles and discrete rotations of the coordinate axes, is
\begin{equation}
  \label{eq:mini3}
  \mat{Z}_1 = a \left(-\frac{1}{2}, -\frac{1}{2\sqrt{3}},0\right)~,
  \quad
  \mat{Z}_2=a\left(\frac{1}{2},-\frac{1}{2\sqrt{3}},0\right)~,
  \quad
  \mat{Z}_3=a\left(0,\frac{1}{\sqrt{3}},0\right)~,   
\end{equation}
with the inertia tensor $ma^2/2\,\mathrm{diag}(1,1,2)$.

The Eckart gauge is defined by (\ref{eq:gge}) with $\Gamma_{a\alpha
  i}=\varepsilon_{aji} Z_{\alpha j}$, $a=1,2,3$, which are normalized
to $\R_1^2 = ma^2/2 = \R_2^2$, $\R_3^2 =ma^2$.  Those gauge conditions
make the planar nature of the problem apparent, since they imply that
$\delta R_{\alpha 3} = 0 = P_{\alpha 3}$, $\alpha=1,2,3$, as
operators.  This leads, in particular, to the operator \mat{\Lambda}\ 
of (\ref{eq:ptrans}) having two null components $\Lambda_1 = 0 =
\Lambda_2$ (as operators), with $\Lambda_3$ being conserved and having
integer eigenvalues, as shown below.  Thus, in this example
\mat{\Lambda}\ is an angular momentum operator, though two- rather
than three-dimensional, and can be rightfully termed ``residual''
angular momentum as in \cite{bou}.  The three- dimensional rotations
of the system are taken into account by the total angular momentum
operator \mat{s}.

Setting $V^{\prime\prime}(a)\equiv m\omega^2$, the quadratic terms in
an expansion of \V\ about $\{\mat{Z}_\alpha\}$ are,
\begin{equation}
  \label{eq:quadv}
  \V_{(2)} = \frac{m\omega^2}{2} \sum_{\alpha<\beta=1}^3
  \left( \frac{1}{a} (\mat{Z}_\alpha - \mat{Z}_\beta)\cdot
  (\mat{\delta R}_\alpha - \mat{\delta R}_\beta) \right)^2~.
\end{equation}
Using the gauge conditions we could eliminate six degrees of freedom,
describing the system in terms of, e.g., $\delta R_{1X}$, $\delta
R_{2X}$, $\delta R_{2Y}$, and their conjugate momenta.  A better
approach is to use a set of normal coordinates $\{\delta
Q_a\}_{a=4}^6$ as discussed in sections \ref{sec:qham} and
\ref{sec:qham2}.  Thus, with $\Gamma_{a\alpha i}$, $a=1,2,3$, as
defined above and $\Gamma_{a\alpha i}$, $a=7,8,9$, as defined by
(\ref{eq:gammahigh}), we can choose $\Gamma_{a\alpha i}$, $a=4,5,6$,
to be those eigenvectors of the quadratic form associated with
$\V_{(2)}$ which satisfy the orthogonality conditions
(\ref{eq:orthocomp}) (in particular, (\ref{eq:orthocomp2})), and
normalized\footnote{In this section and the following we restore
  $\hbar$ in all expressions.} to $\R^2=\hbar/\omega$.  Those
$\Gamma_{a\alpha i}$, $a=4,5,6$, are the vibrational normal modes of
$\V_{(2)}$, whose associated normal coordinates $\delta Q_a$ are given
by (\ref{eq:Q2}),
\begin{equation}
  \label{eq:dQ}
  \begin{aligned}
    \delta Q_4 &= \sqrt{\frac{m\omega}{\hbar}} \left(
      \frac{1}{2}\delta R_{1X} - \frac{1}{2\sqrt{3}}\delta R_{1Y} -
      \frac{1}{2} \delta R_{2X} - \frac{1}{2\sqrt{3}} \delta R_{2Y} +
      \frac{1}{\sqrt{3}} \delta R_{3Y}\right)~,\\
    \delta Q_5 &= \sqrt{\frac{m\omega}{\hbar}} \left(
      -\frac{1}{2\sqrt{3}}\delta R_{1X} - \frac{1}{2}\delta R_{1Y} -
      \frac{1}{2\sqrt{3}} \delta R_{2X} + \frac{1}{2} \delta R_{2Y} +
      \frac{1}{\sqrt{3}} \delta R_{3X}\right)~,\\
    \delta Q_6 &= \sqrt{\frac{m\omega}{\hbar}} \left(
      -\frac{1}{2}\delta R_{1X} - \frac{1}{2\sqrt{3}}\delta R_{1Y} +
      \frac{1}{2} \delta R_{2X} - \frac{1}{2\sqrt{3}} \delta R_{2Y} +
      \frac{1}{\sqrt{3}} \delta R_{3Y}\right)~.
  \end{aligned}
\end{equation}
Similarly, from (\ref{eq:chainQ}), (\ref{eq:PQ}) (with $3N-3$ instead
of $3N$) we obtain the relation between $\partial/\partial\delta Q_a$
and either $\partial/\partial\delta R_{\alpha i}$ or $P_{\alpha i}$.
The result is given by (\ref{eq:dQ}) with $\delta Q_a$ substituted by
$1/i \partial/\partial\delta Q_a$ on the l.h.s., and $\delta R_{\alpha
  i}$ substituted by either $\hbar/(m\omega) 1/i \partial/\partial
R_{\alpha i}$ or $1/(m\omega) P_{\alpha i}$, respectively, on the
r.h.s.

The quadratic piece of the Hamiltonian \Hntr\ (henceforth $\wth{\H}$)
of section \ref{sec:constraint2} can be written as, 
\begin{equation}
  \label{eq:quadham}
  \wth{\H}_{0} \equiv \frac{1}{2m} \sum_{\beta=1}^3 \mat{P}_\beta^2
  + 
  \V_{(2)} =
  \frac{\hbar\omega}{2} \sum_{a=4}^6  \left(
    - \frac{\partial^2}{\partial\delta Q_a^2}
    + \sigma_a^2 \delta Q_a^2 \right)~,
  \qquad
  \sigma_4^2=\frac{3}{2}=\sigma_5^2~,
  \quad
  \sigma_6^2=3~.
\end{equation}
$\wth{\H}_{0}$ is the lowest-order Hamiltonian in a perturbative
expansion in powers of $\epsilon = \sqrt{\hbar/(m\omega a^2)}\ll 1$. 
{F}rom the definition (\ref{eq:ptrans}), or equivalently from
(\ref{eq:residual}), we find the residual angular momentum as,
\begin{equation}
  \label{eq:residualQ}
  \Lambda_i = \sum_{a=4}^{3N-3} \frac{\delta \Q_{ai}}{\R^2}
  \frac{1}{i} \frac{\partial}{\partial\delta Q_a}
  \quad
  \text{and therefore}
  \quad
  \Lambda_1 = 0 = \Lambda_2~,
  \quad
  \Lambda_3 = \frac{1}{i} \left( \delta Q_5
  \frac{\partial}{\partial\delta Q_4} - \delta Q_4
  \frac{\partial}{\partial\delta Q_5} \right)~,
\end{equation}
with $\delta \Q_{ai}$  defined in (\ref{eq:QaiE}).
To $\O(\epsilon^2)$ the quantities entering $\wth{\H}$ are $\R_1^2 =
\R_2^2 = \R_3^2/2 = \hbar/(2\omega\epsilon^2)$, $\N_{ij}^{-1} =
1/\R_{(i)}^2 \delta_{(i)j} + \O(\epsilon^3)$, $\V_1 =
\O(\epsilon^3)$, and the anharmonic terms in \V, which are
$\O(\epsilon^3)$.  The expansion of $\V_2= \mathrm{cst.} +
\O(\epsilon^3)$ starts at $\O(\epsilon^2)$ but the lowest-order term
is a constant, which we drop.  Retaining only terms through
$\O(\epsilon^2)$ in $\wth{\H}$, and expressing them in terms of normal 
coordinates, we obtain
\begin{equation}
  \label{eq:pertham}
  \wth{\H}=\wth{\H}_0 + \wth{\H}_1~,
  \qquad
  \wth{\H}_1 = \frac{\omega \epsilon^2}{\hbar} \left(
    \mat{s}^2 - s_3^2 + \frac{1}{2} (s_3+\Lambda_3)^2\right) + 
    \O(\epsilon^3) ~.
\end{equation}
where we used $\Lambda_3$ as defined in (\ref{eq:residualQ}), and
dropped all constant terms.  The operators $\mat{s}^2$, $s_3$ and
$\Lambda_3$ all commute with each other and with the Hamiltonian.  The
physical meaning of the Hamiltonian (\ref{eq:pertham}) is apparent
from (\ref{eq:ipL}): for physical wave functions the operator
$\mat{s}$ gives the matrix elements of $-\mat{L}$ so that, to this
order, (\ref{eq:pertham}) corresponds to $\wt{\H}_1 = 1/2 \N^{-1}_{ij}
(L_i - \Lambda_i)(L_j - \Lambda_j)$, with $\N^{-1}_{ij}$ the inverse
of the equilibrium inertia tensor and $\mat{L} - \mat{\Lambda}$ the
difference of the total and residual angular momenta in the Eckart
frame.

Since we are not going to compute beyond $\O(\epsilon^2)$ in
perturbation theory, it is easier (to this order) to solve the
eigenvalue problem for (\ref{eq:pertham}) exactly rather than as a
perturbation about the Hamiltonian (\ref{eq:quadham}).  We introduce
cylindrical coordinates in the space of $\delta Q_{4,5,6}$
\begin{equation}
  \label{eq:cylin}
  \rho = \sqrt{\delta Q_4^2 + \delta Q_5^2}~,
  \quad
  \varphi = \mathrm{arctan}\left(\frac{\delta Q_4}{\delta
  Q_5}\right)~, 
  \quad
  \zeta=\delta Q_6~,
\end{equation}
and classify the Hamiltonian eigenfunctions and eigenvalues according
to the eigenvalues of 
\raisebox{0ex}[0ex]{$\wth{\H}_0$},
$\Lambda_3$, $\mat{s}^2$ and $s_3$ (with quantum numbers denoted by
$(n,n_\zeta)$, $\lambda$, $\ell$, $m$, resp.).  The wave functions
are,
\begin{equation}
  \label{eq:waves}
\begin{gathered}
  \wth{\psi}{}_{n\lambda n_\zeta}^{\ell m}
  (\rho,\varphi,\zeta;\ver{e})= 
  R_{n|\lambda|}(\rho) \Phi_\lambda(\varphi) Z_{n_\zeta}(\zeta)
  Y_{\ell m}(\ver{e})\\
  R_{n|\lambda|}(\rho) = \rho^{|\lambda|} L_n^{|\lambda|} \left(
    \sqrt{\frac{3}{2}} \rho^2 \right) e^{-\sqrt{3/2} \rho^2/2}~,
\quad
  \Phi_\lambda(\varphi) = e^{i\lambda\varphi}~,
\quad
  Z_{n_\zeta}(\zeta) = H_{n_\zeta}(\sqrt{3}\zeta) e^{-3\zeta^2/2} ~,
\end{gathered}
\end{equation}
where we omitted a normalization constant, $n$, $n_\zeta$ and $\ell$
are non-negative integers, $\lambda$ and $m$ are integers, and the
spherical harmonics $Y_{\ell m}$, associated Laguerre polynomials
$L_n^k$ and Hermite polynomials $H_n$ are defined in the standard way
in quantum mechanics \cite{gal}.  The dependence of the wave function
(\ref{eq:waves}) on \ver{e}\ only carries the representation of the
rotation group appropriate to a state of angular momentum $\ell$.  We
could as well suppress the dependence on \ver{e}\ and define the wave
function to be a column with $2\ell+1$ components, depending only
on the three vibrational variables $\rho$, $\varphi$, $\zeta$. The
energy eigenvalues are,
\begin{equation}
  \label{eq:energies}
  \begin{gathered}
E_{n\lambda n_\zeta}^{\ell m} = E_{n\lambda n_\zeta}^{(0)} +
  E_{\lambda \ell m}^{(1)} \\
  E_{n\lambda n_\zeta}^{(0)} = \hbar\omega \left(
    \sqrt{3} \left(n_\zeta + \frac{1}{2}\right) + \sqrt{\frac{3}{2}}
    (2n + |\lambda|+1) \right)~,
  \quad
  E_{\lambda \ell m}^{(1)} = \hbar\omega\epsilon^2 \left(\ell (\ell+1)
    - m^2 + \frac{1}{2} (m+\lambda)^2 \right)~.
  \end{gathered}  
\end{equation}
In the vibrational ground state $n=n_\zeta=\lambda=0$,
(\ref{eq:energies}) reduces to the spectrum of an axis-symmetric top. 

\subsection{A simple example with \mat{N=4}}
\label{sec:simple4}

Adding one more particle of the same mass to the model of the previous
section we obtain a system whose classical equilibria are those
configurations with the particles lying at relative rest on the
vertices of a regular tetrahedron of side $a$.  The inertia tensor of
the equilibrium configuration is now $ma^2\mathrm{diag}(1,1,1)$.  The
constants $\Gamma_{a\alpha i}=\varepsilon_{aji} Z_{\alpha j}$,
$a=1,2,3$, defining the Eckart gauge are then normalized to
$\R^2_{1,2,3}=ma^2$.

This system, unlike that of the previous section, is fully
three-dimensional.  With $\epsilon = \sqrt{\hbar/(m\omega a^2)}$, the
$\O(\epsilon^2)$ perturbation $\wth{\H}_1$ does not commute with the
zeroth-order quadratic Hamiltonian, so the $\O(\epsilon^2)$
corrections to the unperturbed energies must be computed
perturbatively.  Diagonalizing $\wth{\H}_1$ within eigenspaces of
$\wth{\H}_0$ is best done numerically, due to the large accidental
degeneracies of the unperturbed levels beyond the ground state.  For
that reason, we will restrict ourselves to making only some remarks on
the form of the $\O(\epsilon^2)$ Hamiltonian.

We have six vibrational normal modes, with normal coordinates
$\{\delta Q_a\}_{a=4}^9$ whose expressions in terms of the position
vectors $\{\mat{\delta R}_\alpha\}$ we omit for brevity.  The
unperturbed Hamiltonian is given by (\ref{eq:quadham}), with $\beta$
now running up to 4, $a$ up to 9, and with $\sigma_4^2=\sigma_5^2=1$,
$\sigma_6^2=\sigma_7^2 =\sigma_8^2=2$ and $\sigma_9^2=4$.  We have
$\N_{ij}^{-1} = 1/(ma^2)\delta_{ij} + \O(\epsilon^3)$, and the quantum
potentials $\V_{1,2}$ and the anharmonic corrections to $\V_{(2)}$
starting at $\O(\epsilon^3)$ up to constant terms.  Thus, to
$\O(\epsilon^2)$ the perturbation Hamiltonian is given by the second
term in (\ref{eq:hamW4}), with the momentum operators of
(\ref{eq:momopT}).  Dropping constant terms, the $\O(\epsilon^2)$
perturbation can be arranged in the form,
\begin{equation}
  \label{eq:pertham4}
  \wth{\H}_1 = \frac{\omega \epsilon^2}{\hbar} \left(
    \mat{s} +\mat{\Lambda}\right)^2 + \O(\epsilon^3) ~,
  \qquad
  \Lambda_i = \frac{\hbar}{i} \sum_{c=4}^9 \frac{1}{\R^2} \delta
    \Q_{ci} \frac{\partial}{\partial \delta Q_c}~.
\end{equation}
The normal coordinate we call $\delta Q_9$ corresponds to a
vibrational mode $\mat{\Gamma}_9$ with $\Gamma_{9\alpha i} \propto
Z_{\alpha i}$, i.e., a dilatation mode.  Explicit computation shows
that $\delta\Q_{9i}=0$, and then $[\mat{\Lambda}, \delta Q_9] =0$.  We
obtain also $[\Lambda_i, \sum_{a=4}^8\delta Q_a^2] =0$, but
$[\Lambda_i, \V_{(2)}] \neq 0$.  The operator \mat{\Lambda}\ is not an
angular momentum operator, as expected on general grounds from
(\ref{eq:comm+b}), but it turns out to be proportional to one,
$[2\Lambda_i, 2\Lambda_j] = i \varepsilon_{ijk} 2\Lambda_k$.  We
cannot give at present necessary and sufficient conditions a many-body
system and a rotating frame must satisfy for \mat{\Lambda}, or a
multiple thereof, to be an angular momentum operator.

\section{Final remarks}
\label{sec:finrem}

The gauge-invariant approach presented here leads to a general and
systematic framework for the quantization of many-body systems in
rotating frames.  Our approach naturally incorporates the notions of
time-dependent symmetry transformations (i.e., gauge transformations),
body-frame time-derivatives (covariant derivatives), moving reference
frames defined as functions of the particle positions (gauge
conditions), and of reference-frame singularities (Gribov ambiguities)
in a most economical way.  It is not, therefore, a superfluous formal
structure imposed on the physics.  The amount of formalism that has
been introduced is in fact minimal.  Rather, we put all those notions
within a consistent mathematical framework.

We have shown that the rotational symmetry of an $N$-body system is a
gauge symmetry, if we restrict ourselves to a fixed angular momentum
sector (eq.\ (\ref{eq:glag})).  Using gauge invariance we formulated
both the classical and quantum theory (in the operator approach) in
rotating frames defined by linear gauge conditions.  In particular, we
explicitly obtained the Hamiltonian operator (\ref{eq:hamS3}) in terms
of position vectors referred to the rotating frame, therefore
constrained by the gauge conditions.  We also showed that the
orientational degrees of freedom can be eliminated from the formalism,
and computed the Hamiltonian operator and inner product within the
corresponding reduced Hilbert space (eqs.\ (\ref{eq:redham}) and
(\ref{eq:ipfinal}), resp.).  A further simplification is obtained by
eliminating the Jacobian from the kinetic energy and the inner
product, leading to the form (\ref{eq:hamW4}) for the Hamiltonian,
including the quantum potentials $\V_{1,2}$ of (\ref{eq:quanpot}), and
(\ref{eq:ipfinal2}) for the inner product.  The Hamiltonian
(\ref{eq:hamW4}), being Weyl-ordered, can be associated to a
generating functional in the path-integral approach with mid-point
discretization.  Those results were extended in section \ref{sec:cmm}
to the translation-invariant case, where the system is further reduced
by eliminating the center-of-mass degrees of freedom.  The particular
case of quasi-rigid systems was discussed in section \ref{sec:quasi}.

The results given in the foregoing apply to a very general class of
models comprising all $N$-particle systems in three-dimensional
Euclidean space with rotation-invariant potentials.  The fact that we
computed the Hamiltonian operator in terms of position vectors
referred to a rotating frame amounts to a purely conventional choice
of coordinates.  Once the Hamiltonian has been given in those
coordinates it is straightforward to transform it to any other
coordinate set, as done in sects.\ \ref{sec:simple3} and
\ref{sec:simple4}, there being no need to compute it again.  For
simplicity, however, we restricted ourselves to systems with
spin-independent interactions.  The extension of the formalism to
include dynamical spin degrees of freedom should in principle be
straightforward.  We note, in this respect, that in order to obtain
half-integer values for the total angular momentum of the $N$-particle
system, when appropriate, we should substitute the rigid rotator in
(\ref{eq:rlag}) by a ``rotator'' with half-integer angular momentum.
We did not consider, either, those cases in which it is not possible,
or desirable, to impose three gauge conditions depending only on
particle coordinates.  Among those are, e.g., one-particle systems
(including translation-invariant two-particle systems), for which the
gauge conditions $R_Y=R_Z=0=\wt{\xi}_X$ lead to a description in
spherical coordinates, and the conditions
$R_Y=0=\wt{\xi}_X=\wt{\xi}_Y$ to cylindrical coordinates.  The case of
reference frames defined by gauge conditions of a more general form
than those considered in the previous sections can also be treated by
the methods discussed in this paper.

\section*{Acknowledgements}

A.O.B.\ would like to thank L.\ N.\ Epele, H.\ Fanchiotti and C.\ A.\  
Garc\'{\i}a Canal for many discussions.  The authors have been
partially supported by Conacyt of Mexico through grant 32598E.

\appendix
\numberwithin{equation}{section}

\section{The Laplacian in configuration space}
\label{sec:appa}

The kinetic energy of the system of particles considered in section
\ref{sec:linear} is proportional to the Laplacian in configuration
space $\nabla^2_q = \sum_{a=1}^{3N} \partial^2/\partial q_a^2$, with
the generalized coordinates $\{q_a\}_{a=1}^{3N}$ defined in
(\ref{eq:q}).  In this appendix we compute the expression for
$\nabla^2_q$ in curvilinear coordinates $\{Q_a\}_{a=1}^{3N}$ defined
as follows.  For $1 \leq a \leq 3$, $Q_a \equiv \theta_a$, with
$\theta_a$ parametrizing \mat{U}\ in (\ref{eq:trans2}), and 
$\{Q_a\}_{a=4}^{3N}$ defined by (\ref{eq:Q}).  The relation between
the two sets of coordinates is given implicitly by (\ref{eq:q}) and
(\ref{eq:Q}).  $\nabla^2_q$ is then given by the standard expression,
\begin{equation}  \label{eq:laplacian}
\begin{gathered}
  \nabla^2_q = \sum_{a,b=1}^{3N} \frac{1}{J} \frac{\partial}{\partial 
    Q_a} M^{-1}_{ab} J \frac{\partial}{\partial Q_b}\\
  M_{ab} \equiv \sum_{a=1}^{3N} \frac{\partial q_c}{\partial Q_a}
  \frac{\partial q_c}{\partial Q_b},
  \quad
  M_{ab}^{-1} = \sum_{a=1}^{3N} \frac{\partial Q_a}{\partial q_c}
  \frac{\partial Q_b}{\partial q_c},
  \quad
  J\equiv \det\left(\frac{\partial q}{\partial Q}\right)
  = \left(\det(\mat{M})\right)^{1/2}~.
\end{gathered}
\end{equation}
In order to obtain an explicit expression for $\nabla^2_q$ we have to
build the matrix $M_{ab}^{-1}$.

For $1 \leq a \leq 3$ we have $\partial Q_a/\partial q_c \equiv
\partial \theta_a/\partial q_c = \sum_{\alpha=1}^N (\partial r_{\alpha
  j}/\partial q_c) (\partial \theta_a/\partial r_{\alpha j})$.  {F}rom
(\ref{eq:lambdas}) we have,
\begin{equation}
  \label{eq:tetar}
  \frac{\partial \theta_a}{\partial r_{\alpha j}} = \frac{1}{2}
  \Lambda_{ia}^{-1} \varepsilon_{mik} \frac{\partial U_{ml}}{\partial
  r_{\alpha j}} U_{kl} = -\Lambda_{ia}^{-1} \sum_{b=1}^3 \Q_{ib}^{-1}
  m_\alpha \Gamma_{b\alpha k} U_{kj}~,
\end{equation}
where in the second equality we used (\ref{eq:Ufund}).  Thus,
\begin{equation}
  \label{eq:tetaq}
  \frac{\partial \theta_a}{\partial q_c} = -\sum_{\alpha=1}^N
  \sum_{b=1}^3 \frac{m_\alpha}{\R_c} \Gamma_{c\alpha j}
  \Gamma_{b\alpha k} \Lambda_{ia}^{-1} \Q_{ib}^{-1} U_{kj}~. 
\end{equation}
For $4 \leq a \leq 3N$ we have, from (\ref{eq:Q}),
\begin{equation}
  \label{eq:Qr}
  \frac{\partial Q_a}{\partial r_{\alpha j}} = \sum_{\beta=1}^N
  \frac{m_\alpha}{\R^2} \Gamma_{a\beta i} \frac{\partial R_{\beta
      i}}{\partial r_{\alpha j}} =
  \frac{m_\alpha}{\R^2} \Gamma_{a\alpha i} U_{ij} - \sum_{d=1}^{3}
  \Q_{ai} \Q_{id}^{-1} \frac{m_\alpha}{\R^2} \Gamma_{d\alpha k}
  U_{kj}~,  
\end{equation}
where the last equality follows directly from (\ref{eq:dR/dr}) and
(\ref{eq:Ufund}).  Notice that, by definition, $\Q_{id}^{-1}$ is a
$3\times 3$ matrix inverse to $\Q_{ai}$ with $1 \leq a \leq 3$, but in
general  $\Q_{ai}\Q_{id}^{-1} \neq \delta_{ad}$ if $4 \leq a \leq 3N$
like in the last term in (\ref{eq:Qr}).  With (\ref{eq:Qr}) and
(\ref{eq:q}) we obtain,
\begin{equation}
  \label{eq:Qq}
  \frac{\partial Q_a}{\partial q_c} = \sum_{\alpha=1}^N
  \frac{m_\alpha}{\R^2} \Gamma_{a\alpha i} U_{ij}
  \frac{\Gamma_{c\alpha j}}{\R_c} - 
  \sum_{\alpha=1}^N \sum_{d=1}^{3} \Q_{ak} \Q_{kd}^{-1}
  \frac{m_\alpha}{\R^2} \Gamma_{d\alpha i} U_{ij}
  \frac{\Gamma_{c\alpha j}}{\R_c}~.
\end{equation}
The matrix elements $M_{ab}^{-1}$ can now be computed, starting with
their definition (\ref{eq:laplacian}), and using the orthogonality and
completeness relations (\ref{eq:orthocomp}), and the definition
(\ref{eq:Qai}) of $\Q_{ai}$.

For $1 \leq a,b \leq 3$ we get
\begin{equation}\label{eq:1313}
  M_{ab}^{-1} = \sum_{c=1}^{3N} \frac{\partial \theta_a}{\partial q_c}
  \frac{\partial \theta_b}{\partial q_c} = \Lambda_{ia}^{-1}
  \Lambda_{jb}^{-1} \sum_{d=1}^3 \R_d^2 \Q_{id}^{-1} \Q_{jd}^{-1} =
  \Lambda_{ia}^{-1} \Lambda_{jb}^{-1} \N_{ij}^{-1}~,
\end{equation}
with $\mat{\N}^{-1}$ defined by the last equality (compare
(\ref{eq:N})).  For $1 \leq a \leq 3$, $4 \leq b \leq 3N$ we get  
\begin{equation}\label{eq:1343}
    M_{ab}^{-1} = \sum_{c=1}^{3N} \frac{\partial \theta_a}{\partial q_c}
    \frac{\partial Q_b}{\partial q_c} = \frac{1}{\R^2}
    \Lambda_{ia}^{-1} \Q_{bj} \N_{ij}^{-1}~.  
\end{equation}
The case $4 \leq a \leq 3N$, $1 \leq b \leq 3$ follows from
(\ref{eq:1343}) by the symmetry of $M_{ab}^{-1}$.  Finally, for $4
\leq a,b \leq 3N$ 
\begin{equation}
  \label{eq:4343}
   M_{ab}^{-1} = \sum_{c=1}^{3N} \frac{\partial Q_a}{\partial q_c}
   \frac{\partial Q_b}{\partial q_c} = \frac{1}{\R^2}
   \delta_{ab} + \frac{1}{\R^4} \Q_{ai} \Q_{bj} \N_{ij}^{-1}~. 
\end{equation}
$M_{ab}^{-1}$ is given by (\ref{eq:1313})--(\ref{eq:4343}) in four
blocks 
$\left(
  \begin{array}{c|c}
  \syl 3\times 3      & \syl 3\times (3N-3)\\\hline
  \syl (3N-3)\times 3 & \syl (3N-3)\times (3N-3)
  \end{array}
\right)$,
\begin{equation}
  \label{eq:matr}
  \begin{aligned}
  \mat{M}^{-1} &= 
  \left( \begin{array}{c|c}
    \mat{\Lambda}^{-1 t} \mat{\N}^{-1} \mat{\Lambda}^{-1} &
    \mat{\Lambda}^{-1 t} \mat{\N}^{-1} \mat{\Q}^{t}/\R^2 \\\hline
    \mat{\Q} \mat{\N}^{-1} \mat{\Lambda}^{-1}/\R^2 &
    \mat{1}/\R^2 + \mat{\Q} \mat{\N}^{-1} \mat{\Q}^{t}/\R^4
    \rule{0pt}{12pt}
         \end{array} \right)\\
    &= 
  \left( \begin{array}{c|c}
    \mat{\Lambda}^{-1 t} & \mat{0} \\\hline
    \mat{\Q}/\R^2        & \mat{\N}^\frac{1}{2}\rule{0pt}{12pt} 
         \end{array} \right)
  \left( \begin{array}{c|c}
    \mat{\N}^{-1}        & \mat{0} \\\hline
    \mat{0}              & \mat{\N}^{-1}/\R^2\rule{0pt}{12pt}
         \end{array} \right)
  \left( \begin{array}{c|c}
    \mat{\Lambda}^{-1}   & \mat{\Q}^t/\R^2 \\\hline
    \mat{0}              & \mat{\N}^\frac{1}{2}\rule{0pt}{12pt} 
         \end{array} \right)~.
  \end{aligned}
\end{equation}
{F}rom the last equality we obtain $J = \R^3 |\Lambda| \J$, with
$|\Lambda| = \det(\mat{\Lambda})$ and $\J=(\det(\mat{\N}))^{1/2}$.
Substituting (\ref{eq:matr}) and $J$ into (\ref{eq:laplacian}) we
obtain $\nabla^2_q = -2 (\Hn - \V)$, with \Hn\ given by
(\ref{eq:hamS1}).

\subsection{Reduced Laplacian and Weyl ordering}
\label{sec:appa1}

In order to eliminate the factors of $J$ from $\nabla^2_q$ in
 (\ref{eq:laplacian}), and to Weyl-order it, we write,
\begin{equation}
  \label{eq:laplacian2}
  \wt{\nabla}^2_q \equiv J^{1/2} \nabla^2_q J^{-1/2} =
  \sum_{a,b=1}^{3N} \left(
    M^{-1}_{ab} \frac{\partial^2}{\partial Q_a\partial Q_b} +
    \frac{\partial M^{-1}_{ab}}{\partial Q_a} \frac{\partial}{\partial
      Q_b} \right)  -
  \frac{1}{J^{1/2}} \sum_{a,b=1}^{3N} \left(\frac{\partial}{\partial
  Q_a}\left(M^{-1}_{ab} \frac{\partial J^{1/2}}{\partial
  Q_b}\right)\right)~. 
\end{equation}
Defining the Weyl-ordered differential operator,
\begin{equation}
  \label{eq:weyldiff}
  \left( M^{-1}_{ab} \frac{\partial}{\partial Q_a}
    \frac{\partial}{\partial Q_b} \right)_W =
  \frac{1}{4} M^{-1}_{ab} \frac{\partial}{\partial Q_a}
    \frac{\partial}{\partial Q_b} + \frac{1}{4}
    \frac{\partial}{\partial Q_a} M^{-1}_{ab} \frac{\partial}{\partial
    Q_b} + \frac{1}{4}
    \frac{\partial}{\partial Q_b} M^{-1}_{ab} \frac{\partial}{\partial
    Q_a} + \frac{1}{4}  \frac{\partial}{\partial Q_a}
    \frac{\partial}{\partial Q_b} M^{-1}_{ab}~,
\end{equation}
(\ref{eq:laplacian2}) becomes,
\begin{equation}
  \label{eq:laplacian3}
  \wt{\nabla}^2_q = 
  \sum_{a,b=1}^{3N} \left( M^{-1}_{ab} \frac{\partial}{\partial Q_a} 
    \frac{\partial}{\partial Q_b} \right)_W -
  \frac{1}{4} \sum_{a,b=1}^{3N} \frac{\partial^2 M^{-1}_{ab}}{\partial
    Q_a \partial Q_b} - \sum_{a,b=1}^{3N} \frac{1}{J^{1/2}} \left(
    \frac{\partial}{\partial Q_a} \left( M^{-1}_{ab} \frac{\partial
  J^{1/2}}{\partial Q_b} \right)\right)~.
\end{equation}
The last two terms on the r.h.s.\ are multiplicative operators which,
up to a constant factor, constitute the quantum potential. The last
one can be considerably simplified by using the second line of
(\ref{eq:laplacian}) to write,
\begin{gather}
  -\sum_{a,b=1}^{3N} \frac{1}{J^{1/2}} \left( \frac{\partial}{\partial
      Q_a} \left(M^{-1}_{ab} \frac{\partial J^{1/2}}{\partial Q_b}
    \right) \right) = \frac{1}{2} \sum_{a,b,c=1}^{3N}
  \frac{1}{J^{1/2}} \frac{\partial}{\partial Q_a} \left(
    \frac{\partial Q_a}{\partial q_c} J^{1/2}
    \left(\frac{\partial}{\partial Q_b} \frac{\partial Q_b}{\partial
        q_c}\right)
  \right)\nonumber\\
  \begin{aligned}  \label{eq:pot1}
    &= \sum_{a,b,c=1}^{3N} \left( \rule{0pt}{20pt} \frac{1}{2}
      \left(\frac{\partial}{\partial Q_a}\frac{\partial Q_a}{\partial
          q_c}\right) \left(\frac{\partial}{\partial Q_b}
        \frac{\partial Q_b}{\partial q_c}\right) + \frac{1}{2}
      \frac{\partial Q_a}{\partial q_c} \left(
        \frac{\partial^2}{\partial Q_a \partial Q_b} \frac{\partial
          Q_b}{\partial q_c}\right) \right.\\
    &\quad + \left.  \frac{1}{4} \frac{\partial Q_a}{\partial q_c}
      \left(\frac{\partial}{\partial Q_b} \frac{\partial Q_b}{\partial
          q_c}\right) \sum_{c^\prime, d=1}^{3N} \frac{\partial
        Q_d}{\partial q_{c^\prime}} \left( \frac{\partial}{\partial
          Q_a} \frac{\partial q_{c^\prime}}{\partial Q_d}\right)
    \right)~,
  \end{aligned}
\end{gather}
where for the last equality we used the analog of (\ref{eq:lemma8}),
\begin{equation}
  \label{eq:lemma8bis}
  \frac{\partial J}{\partial Q_a} = J \sum_{c,d=1}^{3N} \frac{\partial
  Q_d}{\partial q_c} \left( \frac{\partial}{\partial Q_a}\frac{\partial
  q_c}{\partial Q_d}\right)~. 
\end{equation}
The last term in (\ref{eq:pot1}) can be simplified using 
\begin{equation*}
  \sum_{c^\prime, d=1}^{3N} \frac{\partial
    Q_d}{\partial q_{c^\prime}} \left( \frac{\partial}{\partial Q_a}
  \frac{\partial q_{c^\prime}}{\partial Q_d}\right)
  =
  -\sum_{c^\prime, d=1}^{3N} \left( \frac{\partial}{\partial Q_a}
    \frac{\partial Q_d}{\partial q_{c^\prime}} \right)
  \frac{\partial q_{c^\prime}}{\partial Q_d}~,
\end{equation*}
and 
\begin{equation*}
  \sum_{a=1}^{3N} \frac{\partial Q_a}{\partial q_c}
  \left( \frac{\partial}{\partial Q_a}\frac{\partial
      Q_d}{\partial q_{c^\prime}}\right) =
  \sum_{a=1}^{3N} \frac{\partial Q_a}{\partial q_{c^\prime}} 
  \left( \frac{\partial}{\partial Q_a}\frac{\partial
  Q_d}{\partial q_c}\right)~,
\end{equation*}
to obtain,
\begin{equation}
  \label{eq:pot2}
  (\ref{eq:pot1}) = \sum_{a,b,c=1}^{3N} \left( \frac{1}{4}
  \left( \frac{\partial}{\partial Q_a} \frac{\partial Q_a}{\partial
      q_c}\right)
  \left( \frac{\partial}{\partial Q_b} \frac{\partial Q_b}{\partial
      q_c}\right)   +
  \frac{1}{2} \frac{\partial Q_a}{\partial q_c}
  \left( \frac{\partial^2}{\partial Q_a \partial Q_b}
  \frac{\partial Q_b}{\partial q_c}\right)\right)~.
\end{equation}
The second term on the r.h.s.\ of (\ref{eq:laplacian3}) can be
rewritten as,
\begin{align}
  -\frac{1}{4} \sum_{a,b=1}^{3N} \frac{\partial^2
    M^{-1}_{ab}}{\partial Q_a \partial Q_b} &= -\frac{1}{4}
  \sum_{a,b,c=1}^{3N} \left( 2 \left( \frac{\partial^2}{\partial Q_a
        \partial Q_b} \frac{\partial Q_a}{\partial q_c}\right)
    \frac{\partial Q_b}{\partial q_c} + \left(
      \frac{\partial}{\partial Q_b}\frac{\partial Q_a}{\partial
        q_c}\right) \left( \frac{\partial}{\partial Q_a}
      \frac{\partial
        Q_b}{\partial q_c}\right)\right. \nonumber\\
  &\quad \left.  +\left( \frac{\partial}{\partial Q_a}\frac{\partial
        Q_a}{\partial q_c}\right) \left( \frac{\partial}{\partial Q_b}
      \frac{\partial Q_b}{\partial q_c}\right) \right)
  \label{eq:pot3}~.
\end{align}
Thus, finally, substituting (\ref{eq:pot2}) and (\ref{eq:pot3}) into
(\ref{eq:laplacian3}),
\begin{equation}
  \label{eq:laplacianWeyl}
  \wt{\nabla}^2_q = \sum_{a,b=1}^{3N} \left(
    M^{-1}_{ab} \frac{\partial}{\partial Q_a} \frac{\partial}{\partial
      Q_b} \right)_W  -
  \frac{1}{4} \sum_{a,b,c=1}^{3N} \left( \frac{\partial}{\partial Q_b}
  \frac{\partial Q_a}{\partial q_c} \right)
  \left(  \frac{\partial}{\partial Q_a}
  \frac{\partial Q_b}{\partial q_c} \right)~.
\end{equation}

\subsection{The Quantum potential}
\label{sec:quanpot}

The kinetic energy in (\ref{eq:hamW1}) is $-1/2 \wt{\nabla}_q^2$.
Thus, from (\ref{eq:laplacianWeyl}) we have,
\begin{equation}
  \label{eq:quanpot1}
  V_Q = \frac{1}{8} \sum_{a,b,c=1}^{3N} \left(
  \frac{\partial}{\partial Q_b} \frac{\partial Q_a}{\partial
  q_c}\right) \left(
  \frac{\partial}{\partial Q_a} \frac{\partial Q_b}{\partial
  q_c}\right) ~.
\end{equation}
The evaluation of $V_Q$ is straightforward, though rather
laborious.  We closely follow the analogous computation of \cite{chr},
which leads to a compact expression for $V_Q$.  In order to compute
$V_Q$ we split it into three terms,
\begin{equation}
  \label{eq:quanpot2}
  \begin{gathered}
    V_Q = V_{Q_0} + V_{Q_1} + V_{Q_2}~,
    \quad
    V_{Q_0} = \frac{1}{8} \sum_{a,b=4}^{3N} \sum_{c=1}^{3N} \left(
      \frac{\partial}{\partial Q_a} \frac{\partial Q_b}{\partial q_c}
    \right) \left( \frac{\partial}{\partial Q_b} \frac{\partial
        Q_a}{\partial q_c} \right)~,\\
    V_{Q_1} = \frac{1}{4} \sum_{a=1}^{3} \sum_{b=4}^{3N}
    \sum_{c=1}^{3N} \left( \frac{\partial}{\partial \theta_a}
      \frac{\partial Q_b}{\partial q_c} \right) \left(
      \frac{\partial}{\partial Q_b} \frac{\partial
        \theta_a}{\partial q_c} \right)~,
    \quad
    V_{Q_2} = \frac{1}{8} \sum_{a,b=1}^{3} \sum_{c=1}^{3N} \left(
      \frac{\partial}{\partial \theta_a} \frac{\partial
      \theta_b}{\partial q_c} \right) \left( \frac{\partial}{\partial
      \theta_b} \frac{\partial \theta_a}{\partial q_c} \right)~,
  \end{gathered}
\end{equation}
which we consider separately.

\subsubsection{$V_{Q_0}$}

{F}rom (\ref{eq:Qq}) we get $\partial Q_b/\partial q_c$ and thence,
\begin{equation}
  \label{eq:QQq}
\left( \frac{\partial}{\partial Q_a} \frac{\partial Q_b}{\partial q_c}
\right) =
-\sum_{\beta=1}^N \frac{\Gamma_{c\beta n}}{\R_c} \sum_{d=1}^3
\frac{m_\beta}{\R^2} \Gamma_{d\beta s} U_{sn} \left(
  \frac{\partial}{\partial Q_a} \Q_{bg} \Q^{-1}_{gd}
\right)~.
\end{equation}
Substituting this expression, and the corresponding one with $a,b$
interchanged, into $V_{Q_0}$ and using the completeness and
orthogonality relations (\ref{eq:orthocomp}), and (\ref{eq:PQ}),  we
get, 
\begin{equation}
  \label{eq:VQ0a}
  \begin{aligned}
    V_{Q_0} = -\frac{1}{8} \sum_{a,b=4}^{3N} \sum_{\beta,\gamma=1}^N
    \sum_{d=1}^3 \frac{\R_d^2}{\R^4} \Gamma_{a\beta l} \Gamma_{b\gamma
      l^\prime} \left(P_{\beta l} \Q_{bg} \Q^{-1}_{gd}\right)
    \left(P_{\gamma l^\prime} \Q_{am} \Q^{-1}_{md}\right)~.
  \end{aligned}
\end{equation}
In (\ref{eq:VQ0a}) we can extend the summations over $a$ and $b$ down
to 1, due to (\ref{eq:pgauge}).  Expanding the definitions
(\ref{eq:Qai}) of $\Q_{bg}$ and $\Q_{am}$ and using completeness,
(\ref{eq:orthocomp}),
\begin{equation}
  \label{eq:VQ0b}
  V_{Q_0} = -\frac{1}{8} \sum_{\beta,\gamma=1}^{N} \sum_{d=1}^3 \R_d^2
  \varepsilon_{l^\prime gs} \varepsilon_{lmp} \left(P_{\beta l} R_{\gamma
  s} \Q^{-1}_{gd}\right) \left(P_{\gamma l^\prime} R_{\beta
  p} \Q^{-1}_{md}\right)~.
\end{equation}
It is not difficult to check that if in this equation we expand the
expression (\ref{eq:momop}) for $P_{\beta l}$ and $P_{\gamma
  l^\prime}$, the contribution due to the second term in
(\ref{eq:momop}) vanishes, and we get,
\begin{equation}
  \label{eq:VQ0c}
   V_{Q_0} = \frac{1}{8} \sum_{\beta,\gamma=1}^{N} \sum_{d=1}^3 \R_d^2
  \varepsilon_{l^\prime gs} \varepsilon_{lmp}
  \left(\frac{\partial}{\partial R_{\beta l}} R_{\gamma
  s} \Q^{-1}_{gd}\right) \left(\frac{\partial}{\partial R_{\gamma
  l^\prime}} R_{\beta p} \Q^{-1}_{md}\right)~. 
\end{equation}
Using the definition (\ref{eq:Qai}) of $\Q_{ai}$, the derivatives can 
be evaluated to give,
\begin{equation}
  \label{eq:derideri}
  \left(\frac{\partial}{\partial R_{\beta l}} R_{\gamma
      s} \Q^{-1}_{gd}\right) = \left(\delta_{\beta\gamma} \delta_{sl}
      \delta_{gk} - R_{\gamma s} \sum_{a=1}^3 \Q^{-1}_{ga} m_\beta
      \Gamma_{a\beta n} \varepsilon_{nkl}
  \right)\Q^{-1}_{kd}
\end{equation}
and similarly $(\partial/\partial R_{\gamma l^\prime} R_{\beta p}
\Q^{-1}_{md})$.  Thus,
\begin{multline} \label{eq:VQ0d}
  V_{Q_0} = \frac{1}{8} \sum_{\beta,\gamma=1}^{N} \left( \sum_{d=1}^3
    \R_d^2 \Q^{-1}_{kd} \Q^{-1}_{hd} \right) \varepsilon_{l^\prime gs} 
  \varepsilon_{lmp} \left( \delta_{\beta\gamma} \delta_{sl}
    \delta_{gk} - R_{\gamma s} \sum_{a=1}^3 \Q^{-1}_{ga} m_\beta
    \Gamma_{a\beta n}
    \varepsilon_{nkl}\right)\\
  \times \left( \delta_{\beta\gamma} \delta_{pl^\prime} \delta_{mh} -
    R_{\beta p} \sum_{b=1}^3 \Q^{-1}_{mb} m_\gamma \Gamma_{b\gamma
      n^\prime} \varepsilon_{n^\prime hl^\prime}\right)~.
\end{multline}
Identifying the first parenthesis in this expression with
$\N^{-1}_{kh}$ as defined in (\ref{eq:N}), it is straightforward to
rearrange the factors to obtain $V_{Q_0} = \V_2$, with $\V_2$ given in
(\ref{eq:quanpot}).

\subsubsection{$V_{Q_1}$ and $V_{Q_2}$}

With the derivatives $\partial\theta_a/\partial q_c$ given in
(\ref{eq:tetaq}), using the completeness relation (\ref{eq:orthocomp})
we easily get,
\begin{equation}
  \label{eq:VQ2a}
  V_{Q_2} = \frac{1}{8} \sum_{a,b=1}^3 \sum_{\alpha=1}^N
  \sum_{d_1,d_2=1}^3 m_\alpha \Gamma_{d_1\alpha k} \Gamma_{d_2 \alpha
  p} \Q^{-1}_{ld_1} \Q^{-1}_{qd_2}
  \left(\frac{\partial}{\partial\theta_a} \Lambda^{-1}_{qb}
  U_{pj}\right)  
  \left(\frac{\partial}{\partial\theta_b} \Lambda^{-1}_{la}
    U_{kj}\right)~.
\end{equation}
We can write the derivatives $\partial\mat{U}/\partial\theta_a$ in
terms of $\Lambda_{ai}$ using (\ref{eq:lambdas}) to obtain the
expanded form,
\begin{equation}
  \label{eq:VQ2b}
  \begin{aligned}
    V_{Q_2} &= \frac{1}{8} \sum_{a,b=1}^3 \left(\sum_{d=1}^3 \R^2_d
      \Q^{-1}_{ld} \Q^{-1}_{qd} \right)
    \frac{\partial\Lambda^{-1}_{qb}}{\partial\theta_a}
    \frac{\partial\Lambda^{-1}_{la}}{\partial\theta_b}\\
    &\quad +\frac{1}{8} \sum_{a,b=1}^3 \sum_{\alpha=1}^N
    \sum_{d_1,d_2=1}^3 
    m_\alpha \Gamma_{d_1\alpha k} \Gamma_{d_2 \alpha p} \Q^{-1}_{ld_1}
    \Q^{-1}_{qd_2} \varepsilon_{ksp}
    \left(\frac{\partial\Lambda^{-1}_{qb}}{\partial\theta_a} \Lambda^{-1}_{la}
    \Lambda_{bs} - \frac{\partial\Lambda^{-1}_{la}}{\partial\theta_b}
    \Lambda^{-1}_{qb} \Lambda_{as} \right) \\
  &\quad + \frac{1}{8} \sum_{\alpha=1}^N
    \sum_{d_1,d_2=1}^3 m_\alpha \Gamma_{d_1\alpha k} \Gamma_{d_2
    \alpha p} \Q^{-1}_{ld_1} \Q^{-1}_{qd_2} \varepsilon_{plr}
    \varepsilon_{kqr}~. 
  \end{aligned}
\end{equation}
The second line can be evaluated from the commutators
(\ref{eq:commlL}), $[L_i,L_j] = -i \varepsilon_{ijk} L_k$.
Substituting into these commutators the expression (\ref{eq:qpthetaL})
for $L_i$, we are led to
\begin{equation}
  \label{eq:commlamb}
  \sum_{d=1}^3 \left( \Lambda^{-1}_{jd} \frac{\partial
    \Lambda^{-1}_{ic}}{\partial \theta_d} - \Lambda^{-1}_{id}
    \frac{\partial \Lambda^{-1}_{jc}}{\partial \theta_d} \right) =
    \varepsilon_{ijk} \Lambda^{-1}_{kc}~.
\end{equation}
Thus, (\ref{eq:VQ2b}) can be rewritten as,
\begin{equation}
  \label{eq:VQ2c}
  V_{Q_2} = \frac{1}{8} \sum_{a,b=1}^3 \N^{-1}_{lq} 
    \frac{\partial\Lambda^{-1}_{qb}}{\partial\theta_a}
    \frac{\partial\Lambda^{-1}_{la}}{\partial\theta_b}
    +\frac{1}{8} \sum_{\alpha=1}^N \sum_{d_1,d_2=1}^3 m_\alpha
    \Gamma_{d_2 \alpha n} 
    \Q^{-1}_{qd_2} \left( \varepsilon_{nlr} \varepsilon_{kqr} + 
    \varepsilon_{qll^\prime} \varepsilon_{kl^\prime n} \right)
    \Q^{-1}_{ld_1} \Gamma_{d_1\alpha k} ~. 
\end{equation}
The factor in parenthesis in the second term equals
$\varepsilon_{nql^\prime} \varepsilon_{kll^\prime}$, but we will
refrain from simplifying it.  After renaming dummy indices we can
rewrite (\ref{eq:VQ2c}) as,
\begin{equation}
  \label{eq:VQ2d}
    V_{Q_2} = \frac{1}{8} \sum_{a,a'=1}^3 \N^{-1}_{ll^\prime} 
    \frac{\partial\Lambda^{-1}_{l^\prime a^\prime}}{\partial\theta_a} 
    \frac{\partial\Lambda^{-1}_{la}}{\partial\theta_{a^\prime}}
    +\frac{1}{8} \sum_{\alpha=1}^N m_\alpha \sum_{d_1,d_2=1}^3 \left( 
    \Q^{-1}_{l^\prime d_2} \Gamma_{d_2\alpha q} \varepsilon_{qln} -
    \varepsilon_{l^\prime lq} \Q^{-1}_{qd_2} \Gamma_{d_2\alpha n}
    \right) \Q^{-1}_{ld_1} \Gamma_{d_1\alpha k} \varepsilon_{kl^\prime
    n}~, 
\end{equation}
which is exactly analogous to (6.8) of \cite{chr}.

With the derivatives (\ref{eq:tetaq}) and (\ref{eq:Qq}), and using
completeness (\ref{eq:orthocomp}), we can write
\begin{equation}
  \label{eq:VQ1a}
  V_{Q_1} = -\frac{1}{4} \sum_{a=1}^3 \sum_{b=4}^{3N}
  \sum_{\alpha=1}^N \frac{m_\alpha}{\R^2} \left(\Gamma_{b\alpha r} -
  \sum_{d_2=1}^3 
  \Q_{bq} \Q^{-1}_{qd_2} \Gamma_{d_2\alpha r}\right) \frac{\partial
  U_{rl}}{\partial\theta_a} \sum_{d_1=1}^3 \Gamma_{d_1\alpha n}
  \Lambda^{-1}_{l^\prime a} U_{nl} \frac{\partial \Q^{-1}_{l^\prime
  d_1}}{\partial Q_b}~. 
\end{equation}
The derivative $\partial U_{rl}/\partial\theta_a$ can be written in
terms of $\Lambda_{ai}$ with (\ref{eq:lambdas}).  On the other hand,
using (\ref{eq:chainQ}) to write $\partial \Q^{-1}_{l^\prime
  d_1}/\partial Q_b$ in terms of $\partial \Q^{-1}_{l^\prime
  d_1}/\partial R_{\beta l}$ and 
evaluating the latter from (\ref{eq:Qai}), we arrive at
\begin{equation}
  \label{eq:VQ1b}
    V_{Q_1} = -\frac{1}{4} \sum_{\alpha,\beta=1}^N \left(
      \sum_{c,d_1=1}^3 \Q^{-1}_{l^\prime c} \Gamma_{c\beta p}
      \varepsilon_{plm}\right) \left( 
    \Q^{-1}_{md_1} \Gamma_{d_1\alpha n}\right) \sum_{b=4}^{3N}
      \frac{m_\alpha m_\beta}{\R^2} \Gamma_{b\beta l}
      \left(\Gamma_{b\alpha r} - 
      \sum_{d_2=1}^3 \Q_{bq} \Q^{-1}_{qd_2} \Gamma_{d_2\alpha
      r}\right) \varepsilon_{rl^\prime n}~. 
\end{equation}
Expanding $\Q_{bq}$ and using completeness to evaluate
the sum over $b$, with $\D_{lq}^\beta$ defined in (\ref{eq:Dnotat}) we
get 
\begin{equation}
  \label{eq:VQ1c}
    V_{Q_1} = -\frac{1}{4} \sum_{\alpha,\beta=1}^N m_\alpha 
      \sum_{c,d_1=1}^3 \left( \Q^{-1}_{l^\prime c} \Gamma_{c\beta p} 
      \varepsilon_{plm}\right) 
        \left( \Q^{-1}_{md_1}
      \Gamma_{d_1\alpha n}\right) \left( \delta_{\alpha\beta}
      \delta_{lr} - \sum_{d_2=1}^3 m_\beta \D_{lq}^\beta
      \Q^{-1}_{qd_2} \Gamma_{d_2\alpha r}\right)
      \varepsilon_{rl^\prime n}~,
  \end{equation}
an expression which is exactly analogous to eq.\ (6.10) of
\cite{chr}. Combining this last expression for $V_{Q1}$ and
(\ref{eq:VQ2c}) for $V_{Q2}$, we can write,
\begin{equation}
  \label{eq:VQ12}
  \begin{aligned}
    V_{Q_1} + V_{Q_2} &= 
    \frac{1}{8} \sum_{a,a'=1}^3 \N^{-1}_{ll^\prime} 
    \frac{\partial\Lambda^{-1}_{l^\prime a^\prime}}{\partial\theta_a} 
    \frac{\partial\Lambda^{-1}_{la}}{\partial\theta_{a^\prime}}
    -\frac{1}{8} \sum_{\alpha=1}^N m_\alpha \sum_{c,d=1}^3 
    \Q^{-1}_{l^\prime c} \Gamma_{c\alpha l} \Q^{-1}_{md}
    \Gamma_{d\alpha n} (\varepsilon_{nl^\prime p} \varepsilon_{plm} + 
    \varepsilon_{nlp} \varepsilon_{pl^\prime m})\\
    &\quad -\frac{1}{4} \sum_{\alpha,\beta=1}^N m_\alpha m_\beta 
      \sum_{c,d_1=1}^3 \Q^{-1}_{l^\prime c} \Gamma_{c\beta p}
      \varepsilon_{plm} \Q^{-1}_{md_1} \Gamma_{d_1\alpha r}
      \varepsilon_{rl^\prime n} \sum_{d_2=1}^3 \D_{lq}^\beta
      \Q^{-1}_{qd_2} \Gamma_{d_2\alpha n}~.
  \end{aligned}
\end{equation}
The summand on the second line of this equation is the product of
$(\varepsilon_{plm} \D_{lq}^\beta)$ times an expression antisymmetric
in $m$ and $q$.  Thus, we can replace $(\varepsilon_{plm}
\D_{lq}^\beta) \rightarrow 1/2 (\varepsilon_{plm} \D_{lq}^\beta -
\varepsilon_{plq} \D_{lm}^\beta) = 1/2 \varepsilon_{qlm}
\D_{lp}^\beta$, and the second line of (\ref{eq:VQ12}) becomes,
\begin{equation}
  \label{eq:becomes}
  -\frac{1}{8} \left( \sum_{\beta=1}^N m_\beta \sum_{c=1}^3
  \Q^{-1}_{l^\prime c} \Gamma_{c\beta p} \D_{lp}^\beta\right)
  \varepsilon_{qlm} \sum_{\alpha=1}^N m_\alpha 
  \sum_{d_1,d_2=1}^3 \Q^{-1}_{md_1}  \Gamma_{d_1\alpha r}
  \varepsilon_{rl^\prime n}
      \Q^{-1}_{qd_2} \Gamma_{d_2\alpha n}~.
\end{equation}
{F}rom the definitions (\ref{eq:Dnotat}) and (\ref{eq:Qai}), we see that
the factor within parentheses in (\ref{eq:becomes}) reduces to
$(-\delta_{ll^\prime})$.  Making that simplification in
(\ref{eq:becomes}) and substituting the result for the second line of
(\ref{eq:VQ12})  we finally get,
\begin{equation}
  \label{eq:VQ12a}
    V_{Q_1} + V_{Q_2} = \frac{1}{8} \sum_{a,a'=1}^3
    \N^{-1}_{ll^\prime}  \frac{\partial\Lambda^{-1}_{l^\prime
    a^\prime}}{\partial\theta_a}
    \frac{\partial\Lambda^{-1}_{la}}{\partial\theta_{a^\prime}} + \V_1  
\end{equation}
with $\V_1$ given by (\ref{eq:quanpot}). 

\section{Affine transformations and covariant derivatives}
\label{sec:affine}

The transformations (\ref{eq:eutr}) do not depend on $\mat{r}_\alpha$
linearly but affinely, i.e., $ (c_1 \mat{r_\alpha} + c_2
\mat{r_\beta})^\prime = c_1 \mat{r_\alpha}^\prime + c_2
\mat{r_\beta}^\prime$ iff $c_1 + c_2 =1$.
Covariant derivatives are defined so that they transform under
time-dependent transformations in the same way as ordinary derivatives
transform under time-indepent transformations.  Thus, we define $D_t
\mat{r}_\alpha$ as in (\ref{eq:covdev}), so that $(D_t
\mat{r}_\alpha)^\prime = \mat{U}D_t \mat{r}_\alpha$, but $  D_t (c_1
\mat{r}_\alpha + c_2 \mat{r}_\beta) = c_1 D_t \mat{r}_\alpha + c_2 D_t
\mat{r}_\beta$ iff $c_1 + c_2 =1$.
Similarly, the rule for the derivative of a vector product is not the
usual one,  $D_t (\mat{r}_\alpha \wedge \mat{r}_\beta) = (D_t
\mat{r}_\alpha) \wedge \mat{r}_\beta + \mat{r}_\alpha \wedge
(D_t\mat{r}_\beta) + (\mat{r}_\alpha - \mat{r}_\beta)\wedge\mat{\rho}
- \mat{\rho}.$  Since $D_t \mat{r}_\alpha$ transforms linearly under
gauge transformations, we define,
\begin{equation}
  \label{eq:covdev3}
  D_t D_t \mat{r}_\alpha \equiv \frac{d}{dt} (D_t \mat{r}_\alpha) - 
  \mat{\xi} (D_t \mat{r}_\alpha)~,
  \quad
  (D_t D_t \mat{r}_\alpha)^\prime = \mat{U}D_t D_t \mat{r}_\alpha~.
\end{equation}
With (\ref{eq:covdev}) and (\ref{eq:covdev3}), the eqs.\  of motion
from Lagrangian (\ref{eq:rhola}) take the form $m_\alpha D_t D_t
\mat{r}_\alpha = -\mat{\nabla}_\alpha\V$ as in (\ref{eq:eqm}).

With the definition (\ref{eq:angmom}) for the angular momentum
\mat{l}\ and the transformation law (\ref{eq:eutr}) we have
$\mat{l}^\prime = \mat{U}\mat{l}+ M \mat{u}\wedge (\mat{U}D_t\vrcm)$,
with \vrcm\ the center-of-mass position vector.  Thus, we define,
\begin{equation}
  \label{eq:dl}
  D_t\mat{l} \equiv \matd{l} - \mat{\xi} \mat{l} - M \mat{\rho} \wedge 
  (D_t\vrcm)~,
\end{equation}
so that $(D_t\mat{l})^\prime = \mat{U}D_t\mat{l} + M \mat{u}\wedge
(\mat{U}D_tD_t\vrcm)$ with $D_tD_t\vrcm$ defined as in
(\ref{eq:covdev3}) and (\ref{eq:covdev}).  The center-of-mass angular
momentum $\vlcm = M \vrcm \wedge D_t\vrcm$ transforms in the same way
as \mat{l}, and its covariant derivative is defined as in
(\ref{eq:dl}).  {F}rom the eqs.\ of motion for \mat{r_\alpha}\ we then 
get, 
\begin{equation}
  \label{eq:dlm}
  D_t\mat{l}=0~,
  \qquad
  D_tD_t\vrcm=0~,
  \qquad
  D_t\vlcm=0~.
\end{equation}
Therefore, $D_t\mat{l}-D_t\vlcm =
(d/dt-\mat{\xi})(\mat{l}-\vlcm)=0$, which is (\ref{eq:affangmom})
and which, together with the antisymmetry of \mat{\xi}, immediately
leads to $d/dt (\mat{l}-\vlcm)^2=0$.  Furthermore,
$(\mat{l}-\vlcm)^\prime = \mat{U}(\mat{l}-\vlcm)$ so that
$(\mat{l}-\vlcm)^2$ is invariant under gauge transformations,
i.e., frame-independent.

\end{document}